\documentclass[twocolumn,prb,showpacs,multicol,amsmath,amssymb]{revtex4-1}
\usepackage{graphicx}
\usepackage{epstopdf}
\usepackage{bm}% bold math
\usepackage{subfigure}
\usepackage{sidecap}

\usepackage{graphicx}
\usepackage{epstopdf}
\usepackage{amssymb}
\usepackage{amsmath}

\usepackage{bm}% bold math
\usepackage{subfigure}
\usepackage{sidecap}
\graphicspath{{Figures/}}

\newcommand{\si}{\sigma}
\newcommand{\al}{\alpha}

\newcommand{\ga}{\gamma}

\newcommand{\la}{\lambda}
\newcommand{\ka}{\kappa}
\newcommand{\vare}{\varepsilon}
\newcommand{\de}{\delta}
\newcommand{\De}{\Delta}
\newcommand{\Ga}{\Gamma}

\newcommand{\bsig}{{\bm\sigma}}

\newcommand{\bphi}{{\bm\phi}}
\newcommand{\bmu}{{\bm\mu}}

\newcommand{\hphi}{\hat{\phi}}

\newcommand{\be}{\begin{equation}}
\newcommand{\ee}{\end{equation}}

\newcommand{\bea}{\begin{eqnarray}}
\newcommand{\eea}{\end{eqnarray}}
\newcommand{\bd}{\begin{displaymath}}
\newcommand{\ed}{\end{displaymath}}
\newcommand{\ba}{\begin{array}}
\newcommand{\ea}{\end{array}}
\newcommand{\bi}{\begin{itemize}}
\newcommand{\ei}{\end{itemize}}
\newcommand{\bc}{\begin{center}}
\newcommand{\ec}{\end{center}}
\newcommand{\bfl}{\begin{flushleft}}
\newcommand{\efl}{\end{flushleft}}
\newcommand{\bfr}{\begin{flushright}}
\newcommand{\efr}{\end{flushright}}

\newcommand{\bl}{\begin{aligned}}
\newcommand{\el}{\end{aligned}}

\newcommand{\hJ}{\hat{J}}

\newcommand{\hchi}{\hat{\chi}}

\newcommand{\tilh}{\tilde{h}}

\newcommand{\tiD}{\tilde{D}}

\newcommand{\fa}{\frac{a}{2}}
\newcommand{\fc}{\frac{c}{2}}

\newcommand{\fs}{\frac{1}{2}}
\newcommand{\fl}{\frac{3}{2}}
\newcommand{\fu}{\frac{5}{2}}
\newcommand{\fN}{\frac{1}{\sqrt{N}}}

\newcommand{\om}{i\omega_n}
\newcommand{\e}{\epsilon}

\newcommand{\YB}{YbB$_{12}$}
\newcommand{\URU}{UR\lowercase{u}$_2$S\lowercase{i}$_2$}

\def\ket#1{\left\vert #1 \right\rangle}
\def\dg{^{\dagger}}

\def\bR{{\bf R}} 
\def\bk{{\bf k}} \def\bK{{\bf K}} \def\bq{{\bf q}} 
\def\bkq{{\bf k}+{\bf q}} \def\bkQ{{\bf k}+{\bf Q}}
 
\def\bQ{{\bf Q}}  
  
 \def\bd{{\bf d}}  \def\bJ{{\bf J}}

\def\dg{\dagger}

\def\bra{\langle}
\def\ket{\rangle}
\def\={\!\!\!&=&\!\!\!}
\def\+{\!\!\!&&\!\!\!+~}
\def\-{\!\!\!&&\!\!\!-~}

\begin{document}
\date{\today}
\title{Collective spin resonance excitation in the gapped itinerant multipole hidden order phase of URu$_2$Si$_2$}
\author{Alireza Akbari$^{1,2}$}
\email{alireza@apctp.org}
\author{Peter Thalmeier$^3$}
\affiliation{$^1$Asia Pacific Center for Theoretical Physics, Pohang, Gyeongbuk 790-784, Korea\\
$^2$Department of Physics, and Max Planck POSTECH Center for Complex Phase Materials, POSTECH,
Pohang 790-784, Korea\\
$^3$Max Planck Institute for the  Chemical Physics of Solids, D-01187 Dresden, Germany}

\begin{abstract}
 An attractive proposal for the  hidden order (HO) in the heavy electron compound \URU~is an itinerant multipole order of high rank. 
 It is due to the pairing of electrons and holes centered on zone center and boundary, respectively in states that have maximally different total angular momentum components. Due to the pairing with commensurate zone boundary ordering vector the translational symmetry is broken and a HO quasiparticle gap opens below the transition temperature $T_{HO}$. Inelastic neutron scattering (INS) has demonstrated that for $T<T_{HO}$ the collective magnetic response is dominated by a sharp spin exciton resonance at the ordering vector \bQ~that is reminiscent of spin exciton modes found inside the gap of unconventional superconductors and Kondo insulators. We use an effective two-orbital tight binding model incorporating the crystalline electric field effect to derive closed expressions for quasiparticle bands reconstructed by the multipolar pairing terms. We show that the magnetic response calculated within that model exhibits the salient features of the resonance found in INS. We also use the calculated dynamical susceptibility to explain the low temperature NMR relaxation rate.
\end{abstract}


\maketitle

\section{Introduction}
\label{sec:intro}
The continuing investigation of 5f heavy fermion compound ($\gamma \simeq 170$ mJ/mol K$^2$) \URU~is mostly motivated by its
exotic low temperature phases. Firstly a much discussed 'hidden order' (HO) state appears below
$T_{HO}=17.5$ K and at considerably  lower temperature $T_c=1.45$ K unconventional chiral d-wave singlet
superconductivity (SC) appears. In the p-T phase diagram superconductivity is embedded within the HO phase. 
We shall not discuss here the long history of that subject but refer to the 
review in Ref.~\onlinecite{mydosh:11}. 

In the early theoretical work on thermodynamical properties the HO phase was interpreted
in terms of localized and crystalline-electric-field (CEF) split 5f states \cite{santini:94} and their localized multipolar degrees of freedom \cite{kuramoto:09}. However, it has subsequently become evident from comparison of ARPES experiments \cite{meng:13,bareille:14} and local density electronic structure calculations \cite{oppeneer:10} that the itinerant 5f electron band picture is a more appropriate starting point. The LDA calculations show that the heavy Fermi surface in \URU~consists mainly of
electron sheets around the $\Gamma$-point and hole sheets around the body centered tetragonal (bct) Z-point (0,0$,\frac{2\pi}{c}$) consisting mostly of $j=5/2$ 5f orbitals.
Due to hybridization and Coulomb interactions  the effective masses of the main Fermi surface (FS) sheets observed in de Haas-van Alphen (dHvA) and  Shubnikov-de Haas  (SdH) experiments are enhanced by a factor $m^*/m_b\simeq 7-8$ as compared to the band masses \cite{ohkuni:99,aoki:12}. The sheets around $\Gamma$ and $Z$  are nested by the commensurate wave
vector $\bQ = (0,0,\frac{2\pi}{c})$ which is also the ordering vector of the HO phase. This translational symmetry breaking is suggested by ARPES features that indicate a downfolding of the Fermi surface from bct to a simple tetragonal (st) structure connected with gap opening in the nesting region \cite{meng:13,bareille:14}.  In addition HO breaks the tetragonal in-plane symmetry from $C_4$ to $C_2$ as observed in torque \cite{okazaki:11} and cyclotron resonance \cite{tonegawa:12} investigations. There is also direct  evidence from x-ray diffraction of a corresponding reduction in lattice symmetry \cite{tonegawa:14,shibauchi:14}. Furthemore analysis of NMR \cite{takagi:12} and $\mu SR$\cite{kawasaki:14} experiments suggest the breaking of time reversal symmety in the HO phase. Because the tetragonal symmetry breaking can only be achieved by a doubly degenerate HO representation \cite{thalmeier:11,thalmeier:14a} the $E_-$ representation is singled out as best candidate for HO. 
 We stress, however that numerous alternative theoretical models of both localized or itinerant character have been proposed for the HO in \URU. A partial list of these models and their symmetries has been compiled in Ref. \onlinecite{ikeda:12} (Supplement) and  Ref.\onlinecite{thalmeier:14a}.\\ 

In the itinerant picture the HO parameter may be constructed as an electron-hole pair condensate of the nested Fermi surface states. Because the antiferroic (AF) -type \bQ~connects two nested sheets  with very different total angular momentum components  \cite{oppeneer:10,ikeda:12}  $j_z=M,M'$ the HO parameter is not of the usual spin- or charge density wave but has a high multipole rank $r=max|M-M'|$ due to their pairing. This has been explicitly shown in an ab-initio based multi-band interaction model \cite{ikeda:12}  where indeed the leading instability is of $E_-$ type with rank r=5 ('dotriacontapole'). Therefore the e-h pairing of states which contain $j_z=\pm5/2$ components is realized leading to the high-rank multipole character of the HO.  Alternatively to using the ab-initio band structure one may devise a simplified effective two-orbital model for the quasiparticle bands \cite{rau:12} which can be used more easily to study the effect of HO and associated gap opening on various physical problems like quasiparticle interference \cite{akbari:14}, thermodynamic\cite{rau:12}, transport quantities and in particular inelastic neutron spectroscopy (INS) which is investigated here. 

The latter has demonstrated that within the quasiparticle charge gap $\Delta_{HO}=4.1$ meV that opens in the HO phase and is confirmed in STM experiments \cite{park:12} a pronounced and well-developed spin resonance evolves \cite{bourdarot:10,bourdarot:14} at  the ordering wave vector $\bQ=(0,0,\frac{2\pi}{c})$ (which is equivalent to $(\frac{2\pi}{a},0,0)$ in the bct BZ).  It has an order-parameter like temperature dependence of resonance frequency and intensity, where the former reaches $\omega_r=1.7$ meV  for $T\ll T_{HO}$ corresponding to $\omega_r/\Delta_{HO}=0.41$. The spin resonance is a common phenomenon in the gap of, e.g. heavy fermion (HF) unconventional superconductors \cite{stock:12,stockert:11,eremin:08,akbari:12a,chi:08,akbari:09a}, but it may also exist in the hybridization or hidden order gap of non-superconducting Kondo compounds \cite{nemkovski:07,friemel:12,akbari:09,akbari:12,riseborough:92} (for a review see Ref.\onlinecite{thalmeier:15}). As in the case of Kondo insulator \YB~ the spin resonance  in \URU~shows some considerable dispersion when moving away from the commensurate \bQ~into the Brillouin zone. There is a further inelastic excitation at an incommensurate wave vector $\bQ_1 =( 1.4\frac{2\pi}{a},0,0)$ known already much earlier \cite{broholm:87,broholm:91} which, though broadened, exists even above $T_{HO}$ and therefore is not directly linked to HO parameter and gap opening. It has been discussed earlier \cite{balatsky:09} and will not be addressed in the present HO context. We focus exclusively on the low energy spin resonance at commensurate \bQ~which appears only below $T_{HO}$ and is clearly associated with HO gap formation because, unlike the excitation at $\bQ_1$ it vanishes for $p>p_c =1.4$ GPa in the AF phase \cite{bourdarot:10}.\\

The aim of our investigation is to understand and simulate the resonance formation within the itinerant multipolar $E_-$(rank 5) HO scenario motivated above and based on the two-orbital model. We show that the properties of the commensurate resonance found in INS can be described and understood  semi-quantitatively within that itinerant model approach for \URU~by calculating the collective magnetic response in RPA approximation based on the four reconstructed quasiparticle bands in the HO phase. We also show that the hitherto not understood NMR relaxation behavior  in the HO phase may be qualitatively reproduced by the present model as the result of opening of a global spin gap.

In Sec.~\ref{sec:model} we recapitulate the basic ingredients of the model and give the closed solution for the four HO quasiparticle bands. The total angular momentum operators in itinerant basis representation are given in Sec.~\ref{sec:moment}. Then we calculate the bare and collective RPA magnetic response functions in Secs.~\ref{sec:baresus} and \ref{sec:RPA}, respectivly which leads to the magnetic excitation spectrum and INS structure functions. In Sec.~\ref{sec:NMR} we briefly discuss interpretation of NMR within this approach and Secs.~\ref{sec:discussion},\ref{sec:summary} give the discussion of numerical results and conclusion.

\section{Two orbital model for hidden order and heavy quasiparticle bands}
\label{sec:model}

The 5f quasiparticle bands in \URU~are the result of hybridzation of narrow 5f bands exhibiting strong Coulomb
interactions with wider conduction bands. Their enhanced effective masses $(m^*=8-25 m_0)$ \cite{ohkuni:99,aoki:12} cannot adequately be described by
band structure calculations $(m_b\leq 8m_0)$ \cite{ohkuni:99} although the latter give a Fermi surface sheet topology in agreement with dHvA and SdH experiments \cite{oppeneer:10,mydosh:11}.
Conceptually mass enhancement can be obtained within the Anderson lattice model in constrained mean field
approximation \cite{hewson:93} which also leads to a narrow indirect hybridization gap close to the Fermi level, equivalent to the Kondo temperature  ($T_K\simeq 11.1$ meV or $129$ K)  that can be
seen in STM experiments \cite{aynajian:10,park:12}. Here we are primarily interested in the very low energy magnetic response of the heavy quasiparticle $\omega\simeq\Delta_{HO} = 4.1$ meV which is considerably smaller than the Kondo temperature $T_K = 2.7 \Delta_{HO}$. 
Therefore we do not start from an Anderson lattice type model with hybridizing narrow 5f and wide conduction
bands. We rather model directly the heavy quasiparticle bands and their Fermi surface by a simple effective tight
binding model. Ab-initio LDA calculations \cite{oppeneer:10,ikeda:12} have shown that states at the Fermi surface consist mostly of jj-coupled
total angular mometum 5f states $|j = \frac{5}{2},M\rangle$ with $M=\pm\frac{3}{2}, \pm\frac{5}{2}$ . Therefore the heavy bands, which have twofold Kramers- pseudo spin degeneracy may be directly described within an effective two-orbital tight binding model due to Rau and Kee \cite{rau:12} The basis states and pseudospins of this model are defined by
%
%%%%%%%%%%%%%%%%%%%%%%%%%%%%%%%%%%%%%%%%%%%%%%%%%%%%%%%%%%%%%%%%%%%%%
\begin{figure*}
\includegraphics[width=0.3\linewidth,clip]{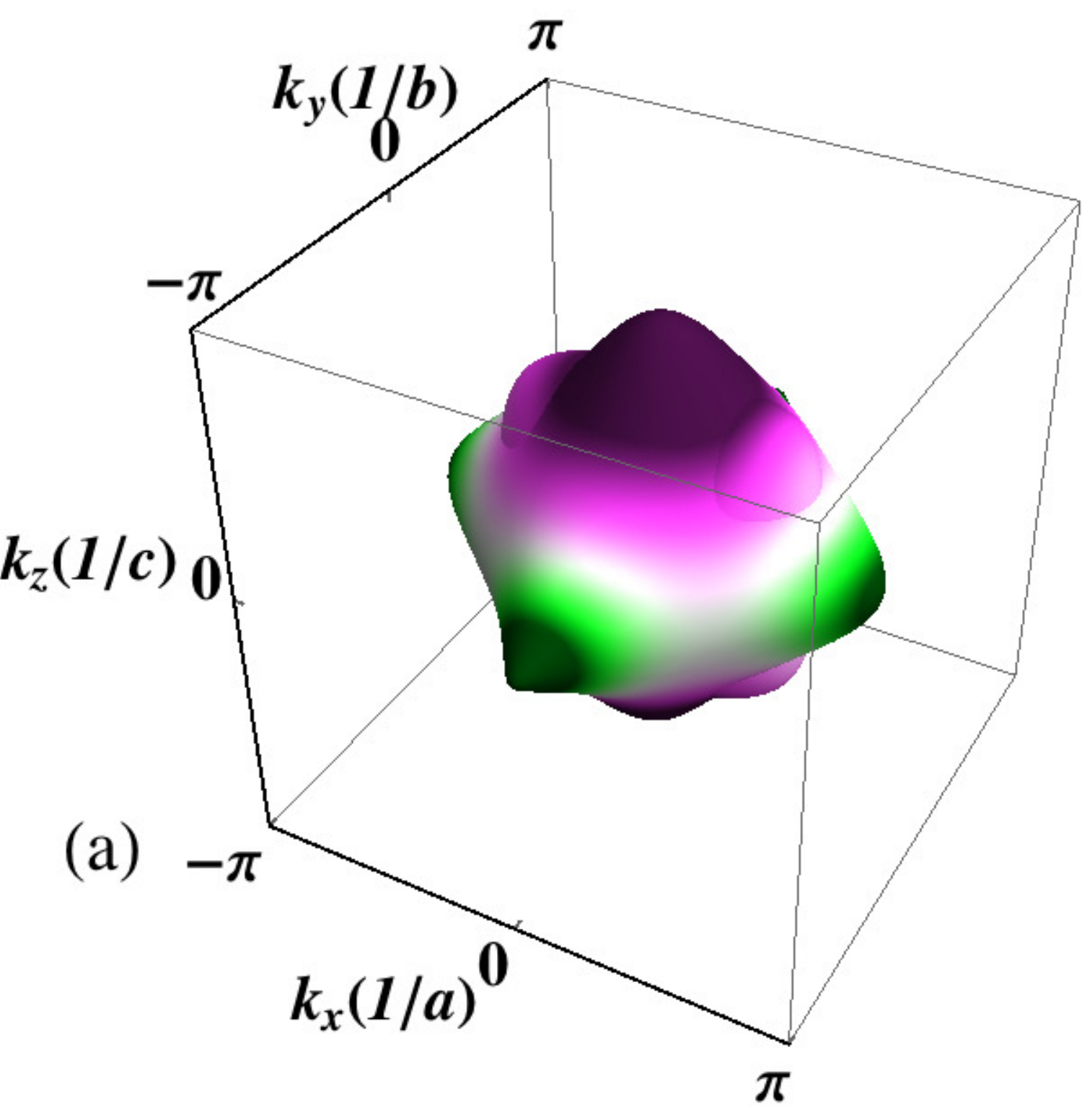}\hspace{0.3cm}
\includegraphics[width=0.28\linewidth,clip]{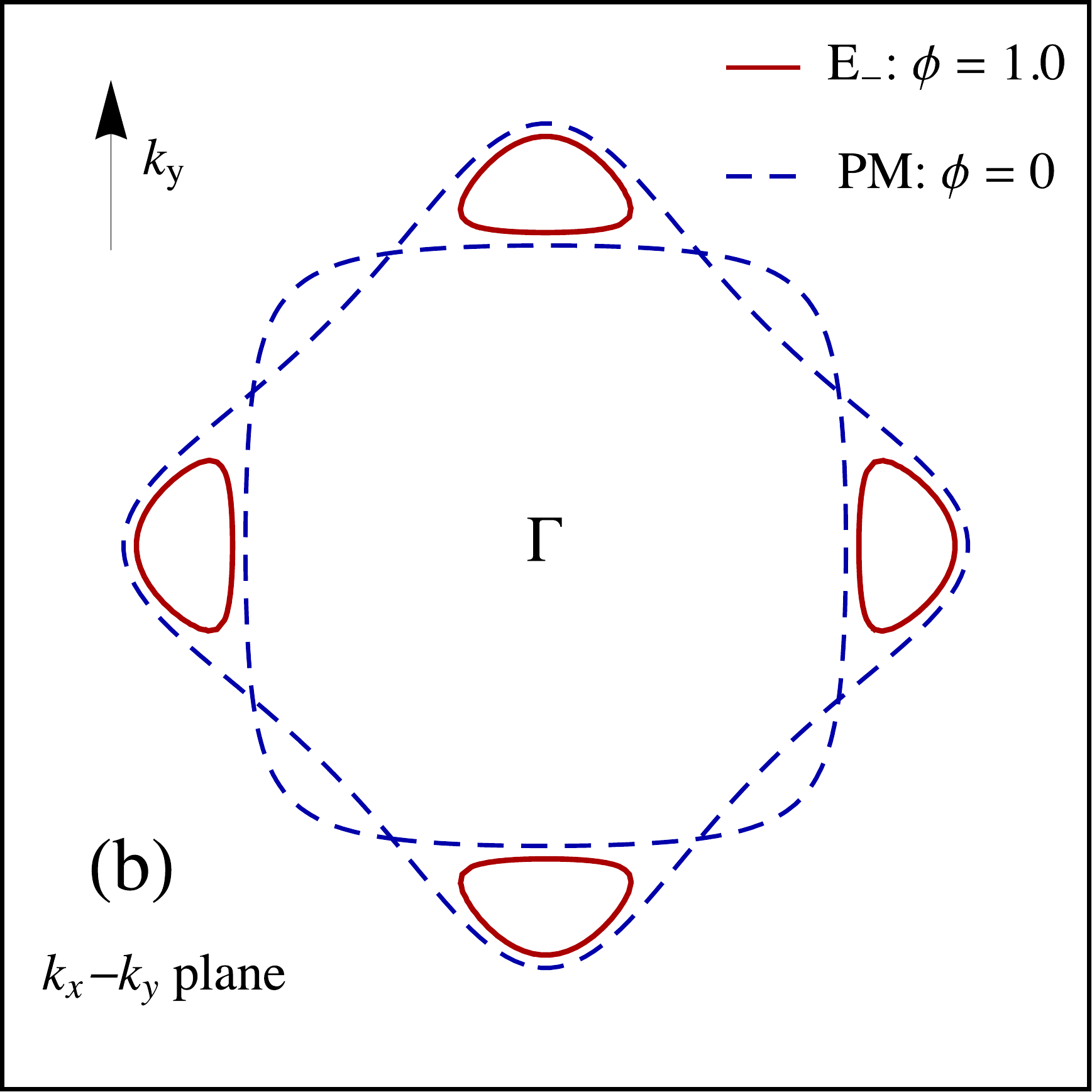}\hspace{0.3cm}
\includegraphics[width=0.3\linewidth,clip]{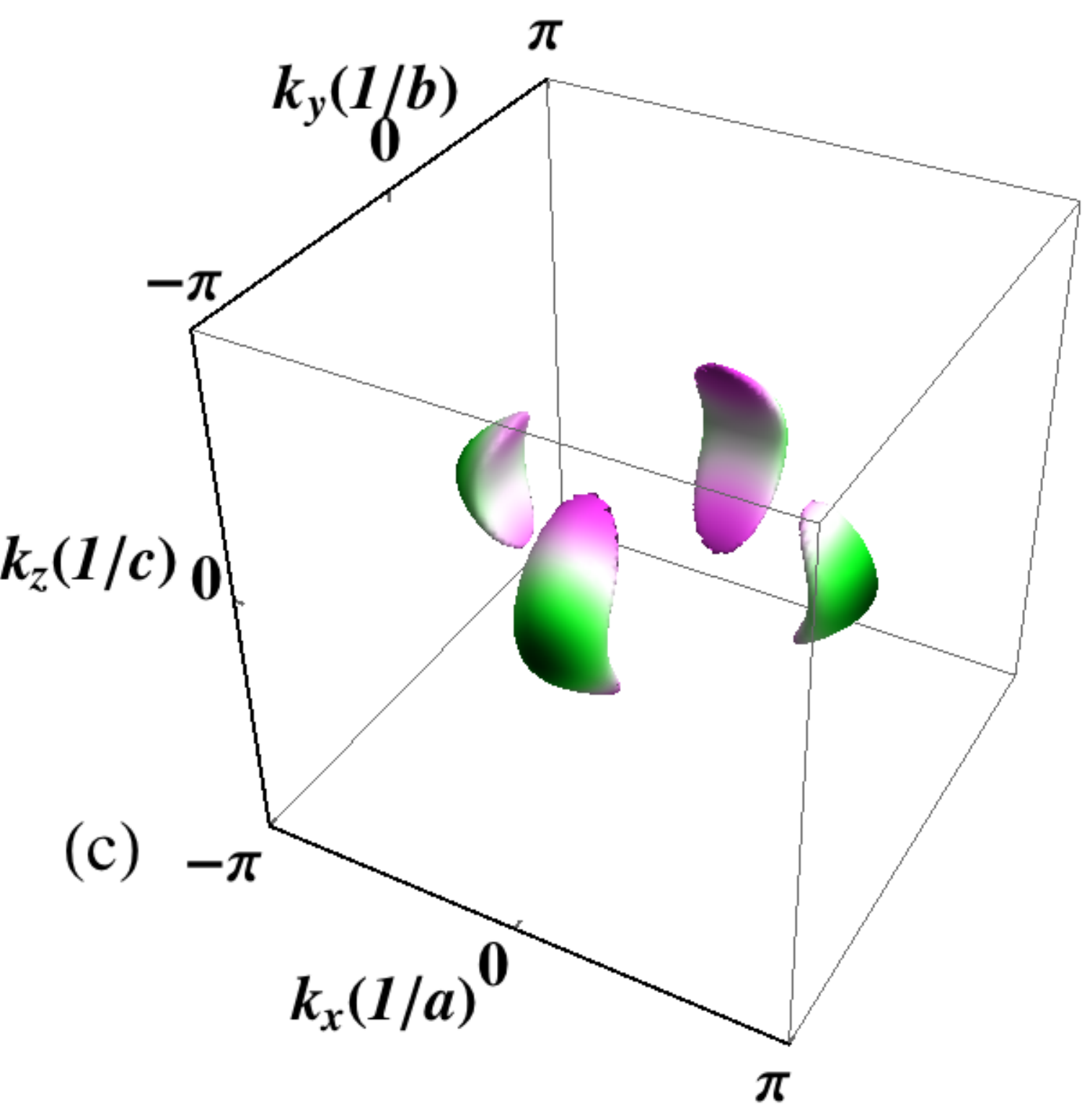}
\caption{
(Color online) 
(a) Fermi surface of the paramagnetic phase downfolded with \bQ~to the st BZ of the HO phase consisting of two closed interpenetrating sheets.
b) FS cuts in the $k_z=0$ plane for paramagnetic (PM) phase ($\phi=0$; blue dashed line) and HO phase ($\phi=1$ with red solid line). 
Momentum range in (b) is given by $-\pi\leq k_{i}\leq \pi$.
(c) Fermi surface sheets of \URU~in the $E_-(1,1)$ HO phase  ($\phi=1$) in the st BZ. The FS is reconstructed in \bk -space regions connected by the nesting vector $\bQ = (0,0,1)$ and breaks up into four smaller sheets without $C_4$ symmetry (for $|k_z|>0$).
%\vspace{0.5cm}
}
\label{fig:Fig1}
\end{figure*}
%%%%%%%%%%%%%%%%%%%%%%%%%%%%%%%%%%%%%%%%%%%%%%%%%%%%%%%%%%%%%%%%%%%%%
%

%
\be
\bl
&
\left(
\begin{array}{c}
f_{1\pm}\\
f_{2\pm}
\end{array}
\right)
=
\left(
 \begin{array}{cc}
\cos\theta& \sin\theta \\
 -\sin\theta& \cos\theta
\end{array}
\right)
\left(
\begin{array}{c}
f_{\pm\frac{5}{2}}\\
f_{\mp\frac{3}{2}}
\end{array}
\right);
%\;\;
%\\
\el
\ee
\be
\bl
&
S^\la_{\al\al'}=\frac{1}{2}\sum_{\si\si'}f^\dg_{\al\si}\si^\la_{\si\si'}f_{\al'\si'}.
\label{eq:CEFtrans}
\el
\ee
where $\la=x,y,z$. Here $\al=1,2$ is the orbital, $\pm$ their Kramers pseudo spin degeneracy index and  $S^\la_{\al\al'}$ is the pseudospin in terms of the orbital creation and annihilation operators \cite{takimoto:08,thalmeier:11,ikeda:12}. The mixing angle $\theta$ results from the CEF potential and is determined by the uniaxial anisotropy of the susceptibility (Sec.~\ref{sec:discussion}). In this basis the kinetic energy may be written as
%\begin{widetext}
%
\be
\bl
H_0=&\sum_{\bk\si}\bigl(A_{1\bk}f^\dg_{1\si\bk}f_{1\si\bk}+A_{2\bk}f^\dg_{2\si\bk}f_{2\si\bk}\bigr)
+
\\
&
\sum_\bk\bigl[D_{\bk}\bigl(f^\dg_{1+\bk}f_{2-\bk}-f^\dg_{2+\bk}f_{1-\bk}\bigr)
%+D^*_{\bk}\bigl(f^\dg_{2-\bk}f_{1+\bk}-f^\dg_{1-\bk}f_{2+\bk}\bigl)
+H.c.
\bigr],
\label{eq:bands}
\el
\ee
%
%\end{widetext}
where the model parameters of Refs.~\onlinecite{rau:12,akbari:14} are implied which are also given in Appendix \ref{sec:app1}. Following Refs.~\onlinecite{ikeda:12, rau:12, akbari:14} the HO is a particle-hole condensate of multipolar rank 5 which is breaking the following tetragonal space group symmetries \cite{shibauchi:14}: i) translational symmetry due to AF structure of HO with $\bQ=(0,0,\frac{2\pi}{c})$ \cite{meng:13,bareille:14}  ii) rotational in-plane symmetry $C_4$ which is reduced to just $C_2$ symmetry \cite{okazaki:11}  iii) time reversal symmetry \cite{takagi:12}. It is
described by a doubly degenerate $E_-$ representation of the order parameter $\bphi=(\phi_x,\phi_y)$. 
Using the pseudospin representation \cite{takimoto:08,thalmeier:11,ikeda:12,thalmeier:14a} ($i=$ site)
the $E_-$ rank-5 'dotriakontapole' is defined by \cite{ikeda:12,rau:12,akbari:14}

\be
\hat{\phi}_x^{E_-}(i)=
\frac{1}{\sqrt{2}}(S_{12}^x+S_{21}^x)_i;\;\;
\hat{\phi}_y^{E_-}(i)=\frac{1}{\sqrt{2}}(S_{12}^y+S_{21}^y)_i.
\label{eq:OP}
\ee
In the HO phase one has to add the molecular field term to $H_0$ ($\kappa=1/(2\sqrt{2})=0.35$):
\be
 H_\phi \!=-\kappa\bphi^\bQ\cdot\sum_\bk(f^\dg_{1\bk}\bsig f^{}_{2\bk+\bQ}+f^\dg_{2\bk}\bsig f^{}_{1\bk+\bQ}) +H.c.
\label{eq:HOP}
\ee
The total mean field Hamiltonian, including the HO molecular fields may be written as
\be
H=H_0+H_\phi=\sum_\bk\Psi^\dg_\bk h_\bk \Psi_\bk;  \;\; h_{\bk}= h_{a\bk}\oplus h_{b\bk}.
\label{eq:HAMfac}
\ee
It consists of two  ($\gamma = a,b$ is the effective Kramers degeneracy index) $4\times 4$ blocks due to combined 
time reversal and translational symmetry $T\otimes t_\bQ$ . Defining the spinor basis 
\be
\bl
\psi^\dg_{a\bk}
&
=(f^\dg_{1+\bk},f^\dg_{2-\bk},f^\dg_{1+\bk+\bQ},f^\dg_{2-\bk+\bQ})
\\
&
=(\psi^\dg_{a\bk 1},\psi^\dg_{a\bk 2},\psi^\dg_{a\bk 3},\psi^\dg_{a\bk 4}),
%;\;\;
%\nonumber
%
\\
\psi^\dg_{b\bk}
&
=(f^\dg_{1-\bk},f^\dg_{2+\bk},f^\dg_{1-\bk+\bQ},f^\dg_{2+\bk+\bQ})
\\
&
=(\psi^\dg_{b\bk 1},\psi^\dg_{b\bk 2},\psi^\dg_{b\bk 3},\psi^\dg_{b\bk 4}),
\label{eq:NambuE}
\el
\ee
and using $\tilh_\bk=(h_{\bk}-\omega I)=\tilh_{a\bk}\oplus\tilh_{b\bk}$ 
the blocks $\tilh_{\ga\bk}$ in corresponding order to Eq.~(\ref{eq:NambuE}) are given by:
\be
\bl
&
h_{a\bk}=
\left(
 \begin{array}{cccc}
A_{1\bk}& D_\bk & 0 & \la_- \\
D_\bk^*& A_{2\bk} & \la_+& 0 \\
0& \la_- & A_{1\bkQ} & D_{\bkQ} \\
\la_+& 0 & D^*_{\bkQ} & A_{2\bkQ}
\end{array}
\right)
;\
%\nonumber
\\
&
h_{b\bk}=
\left(
 \begin{array}{cccc}
A_{1\bk}& -D^*_\bk & 0 & \la_+ \\
-D_\bk& A_{2\bk} & \la_-& 0 \\
0& \la_+ & A_{1\bkQ} & -D^*_{\bkQ} \\
\la_-& 0 & -D_{\bkQ} & A_{2\bkQ}
\end{array}
\right).
\label{eq:hamblock}
\el
\ee
Here $A_{\al\bk}$  ($\al=1,2$ orbital index) and $D_\bk$ are defined in Appendix  \ref{sec:app1}. Furthermore
$\la_\pm=-\kappa(\phi_x\pm i\phi_y)$ with $\kappa=1/2\sqrt{2}=0.35$. We note that $h_{a\bk}\leftrightarrow h_{b\bk}$ under  transformation $D_\bk\rightarrow D^*_{-\bk}= -D^*_{\bk}$ and $\la_\pm\rightarrow\la^*_\pm=\la_\mp$. Although the $E_-$ HO breaks time reversal, the product with translation $t_\bQ$ ($\bk\rightarrow \bk+\bQ$) is still a preserved symmetry. Therefore the four $(\nu=1-4)$ quasiparticle bands obtained from  solving $|h_{\ga\bk}-\omega 1|=0$ still have an additional twofold quasi-Kramers $(\ga =a,b$) degeneracy. They are given by ($\nu=\pm, 1,2; \equiv \nu=1-4$ )  \cite{akbari:14}
\begin{equation}
\bl
&
\vare_{i\bk}=\vare^\pm_{1,2}(\bk)=
%\non
\\
&A_\bk^\perp\pm
\sqrt{ A_\bk^{z2}+\Gamma^2_\bk+\kappa^2|\bphi |^2
\pm 2
\sqrt{A_\bk^{z2}\Gamma^2_\bk+\kappa^2|\bphi |^2\zeta_\bk}
},
\label{eq:dispE}
\el
\end{equation}
where $\Gamma^2_\bk=\Delta_\bk^{\perp 2}+|D_\bk|^2$. On the l.h.s. $\pm$ corresponds to the second (under the square root) and $1,2$ to the first (outside the square root)  $\pm$ on the r.h.s., respectively. Expressed in a simplified way, $\Delta_\bk^\perp$ is the splitting of CEF orbitals 1,2; $A_\bk^z$ is an intra-orbital and $D_\bk$ an inter-orbital $(\sim t_{12})$ hopping energy, respectively (Appendix \ref{sec:app1}).
Furthermore $\phi=|\bphi|=(\phi_x^2+\phi_y^2)^\frac{1}{2}$ is the order parameter amplitude  with $\hat{\bphi}=\bphi/|\bphi|= (\hphi_x,\hphi_y)$. Using $D_\bk=D'_\bk+iD''_\bk$ we introduce the azimuthal function $\zeta_\bk$ which leads to the breaking of fourfold $C_4$ symmetry in the HO phase. It has the general form 
\be
\bl
&
\zeta_\bk
=\frac{1}{2}(|D_\bk|^2-\tiD_\bk)=
\\
&
\frac{1}{2}\bigl[{D'_\bk}^2 +{D''_\bk}^2+(D'_\bk\hphi_y+D''_\bk\hphi_x)^2
- (D'_\bk\hphi_x-D''_\bk\hphi_y)^2\bigr].
\label{eq:zeta}
\el
\ee
Due to $D_{\bk+\bQ}=-D_\bk$ it also satisfies $\zeta_{\bk+\bQ}=\zeta_\bk$.
Here we consider only the $E_-(1,1)$ phase with $\hphi_x=\hphi_y$  and then it simplifies to 
\be
\bl
\zeta_\bk  = \frac{1}{2}(D'_\bk+D''_\bk)^2  
= & 32t_{12}^2\cos^2\fa k_x\sin^2\fa k_y\sin^2\fc k_z ,
\el
\ee
which shows manifestly the fourfold symmetry breaking under $(k_x,k_y)\rightarrow (k_y,-k_x)$ . However, the quasiparticle bands  exhibit the translational symmetry $\vare_{\nu\bk+\bQ}=\vare_{\nu\bk}$ of the HO phase. 

The unitary transformations $U_{\ga\bk}$ that diagonalize $h_{\ga\bk}$ are defined by ($(\nu,\nu'=1-4)$ HO quasiparticle band index and $\ga =a,b$ effective Kramers degeneracy of bands):
\bea
h'_{\ga\bk}&=&U_{\ga\bk}h_{\ga\bk}U^\dagger_{\ga\bk};\;\;\
\{h'_{\ga\bk}\}_{\nu\nu'}=\vare_{\nu\bk}\delta_{\nu\nu'},
\label{eq:hdiagonal}
\eea
Here $\epsilon_{\nu\bk}$ is degenerate in $\gamma=a,b$. The columns of  $U_{\ga\bk}^\dg$ are the eigenvectors of the block matrices  $h_{\ga\bk}$. The corresponding HO quasiparticle operators are defined by
\bea
\psi'_{\ga\bk}=U_{\ga\bk}\psi_{\ga\bk}; \;\;\;  \psi_{\ga\bk}=U^\dg_{\ga\bk}\psi'_{\ga\bk}
.
\label{eq:quasicoord}
\eea
They satisfy canonical commutation relations $\{\psi'_{\nu\ga\bk},\psi'^{\dg}_{\nu'\ga'\bk'}\}=\delta_{\nu\nu'}\delta_{\ga\ga'}\delta_{\bk\bk'}$
and lead to a diagonal representation of the Hamiltonian given by
\be
H=\sum_{\nu\ga\bk}\vare_{\nu\bk}\psi^{'\dg}_{\nu\ga\bk}\psi'_{\nu\ga\bk}.
\ee
%
%%%%%%%%%%%%%%%%%%%%%%%%%%%%%%%%%%%%%%%%%%%%%%%%%%%%%%%%%%%%%%%%%%%%%
\begin{figure}
\includegraphics[width=\linewidth,clip]{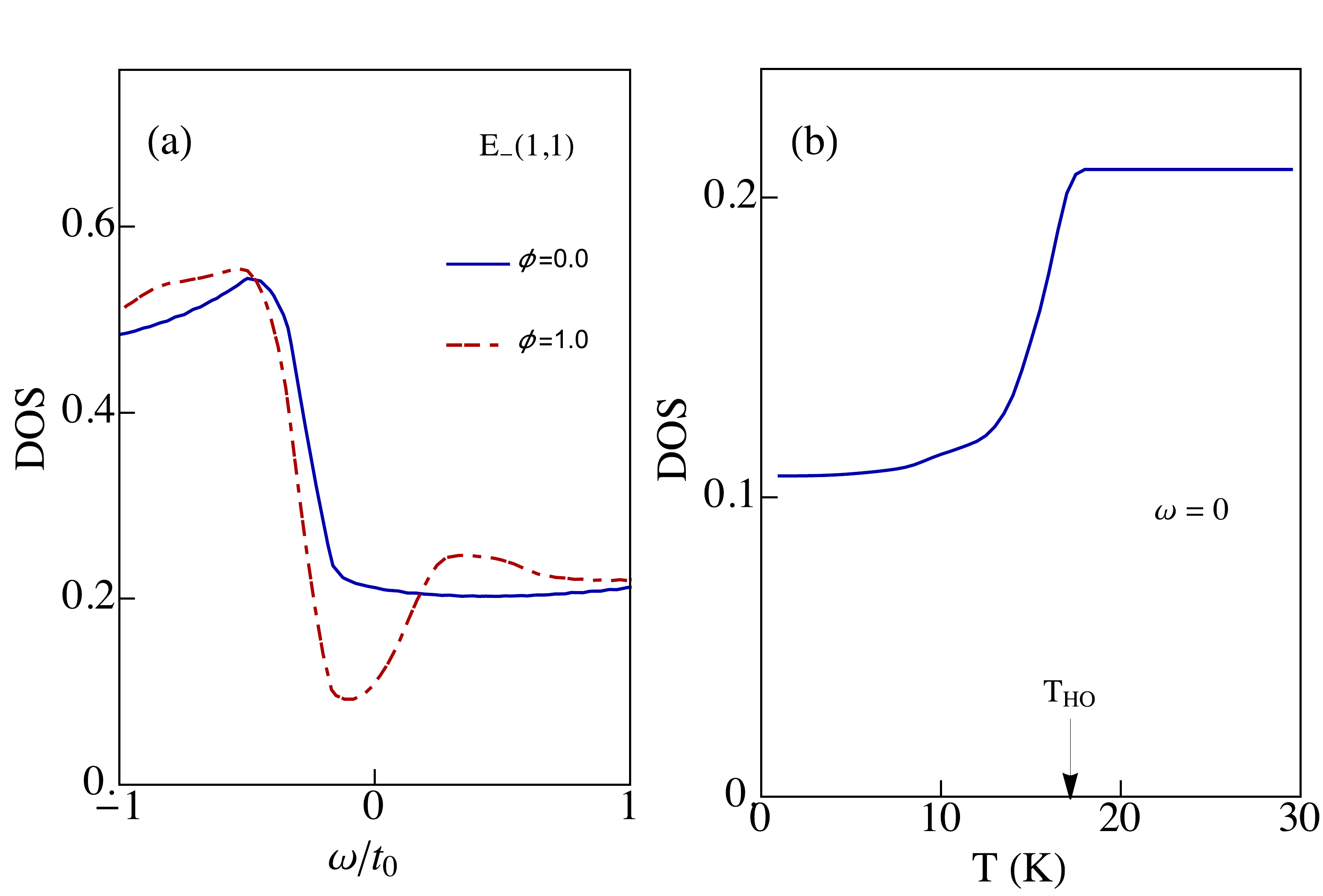}
\caption{
(Color online) 
(a)  Density of states (DOS) showing the evolution of HO gap for small $\omega$. 
(b) Temperature dependence of DOS  at the Fermi energy ($\omega=0$). Here we used a mean field-type interpolation  
$\phi(T)=\phi\tanh(1.76\sqrt{T_{HO}/T-1})$ for the temperature dependence of HO parameter.
%\vspace{0.5cm}
}
\label{fig:Fig2}
\end{figure}
%%%%%%%%%%%%%%%%%%%%%%%%%%%%%%%%%%%%%%%%%%%%%%%%%%%%%%%%%%%%%%%%%%%%%

The Fermi surface of the disordered phase ($\phi=0$) corresponding to the quasiparticle bands in Eq.~(\ref{eq:dispE}) is shown in Fig.~\ref{fig:Fig1}a 
downfolded to the st Brillouin zone of the HO phase. It consists of two interpenetrating electron and hole surfaces whose $k_z=0$ cut is shown in 
 Fig.~\ref{fig:Fig1}b by dashed lines. The points of crossing (in the downfolded or projected picture) of both sheets are connected by the nesting vector
 \bQ~ and therefore when HO appears ($\phi >0$) these regions become gapped and the Fermi surface turns into dissected petal-like shapes shown as full lines in  Fig.~\ref{fig:Fig1}b. This is also visible in the 3D FS presentation of the HO phase in  Fig.~\ref{fig:Fig1}c. It is obvious that large parts of the Fermi surface are removed in the HO phase. These model FS sheets correspond well to the $\beta-$ band sheets from LDA \cite{oppeneer:10} calculations and dHvA and SdH experiments \cite{ohkuni:99,aoki:12} as well as ARPES \cite{meng:13,bareille:14} results. They are also the heaviest sheets with $m_\beta^*\simeq 25 m_0$. From the shrinking of the FS in HO phase one expects that a pronounced suppression of the DOS at the Fermi level $(\omega =0)$ appears. This is shown in Fig.~\ref{fig:Fig2}a in comparison to the disorderd phase. Then, using a mean-field type interpolation for the temperature dependence of the order parameter according to $\phi(T)=\phi\tanh(1.76\sqrt{T_{HO}/T-1})$ the DOS at the Fermi level $(\omega =0)$ is progressively reduced to about one half of the original value (for $\phi=1$) when temperature drops below $T_{HO}$ (Fig.~\ref{fig:Fig2}b).

The breaking up of the large FS sheets into smaller petals and their downfolding to the $\Gamma$ point in the HO phase was found in ARPES experiments \cite{meng:13,bareille:14}. The reduction of the DOS explains the observed entropy loss and suppressed $\gamma$-value in the HO phase \cite{maple:86}. 

\section{The magnetic moment operators in the two-orbital model}
\label{sec:moment}

The dynamical susceptibility which appears in the INS structure factor and cross section contains the total angular 
mometum operator of 5f states. Therefore we need its representation in terms of single particle creation and annihilation 
operators of  $j=\frac{5}{2}$ eigenstates. It is defined by
\be
J_\la=\sum_{MM'}\langle M|J_\lambda|M'\rangle f^\dg_Mf_{M'},
\ee
with the explicit matrix elements given in Appendix \ref{sec:app2}. This leads to
\be
\bl
J^i_x
=&\mu(f^\dg_\fu f_\fl+f^\dg_\fl f_\fu)_i+\mu(f^\dg_{-\fu} f_{-\fl}+f^\dg_{-\fl}f_{_{-\fu}})_i,
%\non
\\
J^i_y
=&-i\mu(f^\dg_\fu f_\fl-f^\dg_\fl f_\fu)_i+i\mu(f^\dg_{-\fu} f_{-\fl}-f^\dg_{-\fl}f_{_{-\fu}})_i,
%\non
\\
J^i_z
=&\fu(f^\dg_\fu f_\fu-f^\dg_{-\fu}f_{-\fu})_i+\fl(f^\dg_{\fl}f_{\fl}-f^\dg_{-\fl}f_{-\fl})_i
.
\label{eq:momentloc1}
\el
\ee
The magnetic moment operators can also be written in pseudo-spin components $S^{i\la}_{\al\al'}$ at site $i$ 
$(\al,\al'=1,2$ orbital index, $\la=x,y,z$ cartesian index) or in terms of their Fourier components.
They are obtained from Eq.~(\ref{eq:CEFtrans}) as
\be
\bl
S^{\bq\la}_{\al\al'}
=&
\fN\sum_ie^{i\bq\bR_i}f^\dg_{\al\si}(i)\si^\la_{\si\si'}f_{\al'\si'}(i)
\\
=&                   
\fN\sum_\bk f^\dg_{\al\si\bk}\si^\la_{\si\si'}f_{\al'\si'\bk+\bq}.
\label{eq:FTdef}
\el
\ee
Using the  transformation of Eq.~(\ref{eq:CEFtrans}), inserting into Eqs.~(\ref{eq:momentloc1}) and performing
the Fourier transform we get \cite{rau:12}:
%
%
%%%%%%%%%%%%%%%%%%%%%%%%%%%%%%%%%%%%%%%%%%%%%%%%%%%%%%%%%%%%%%%%%%%%%
\begin{figure}
\includegraphics[width=1.2\linewidth,clip]{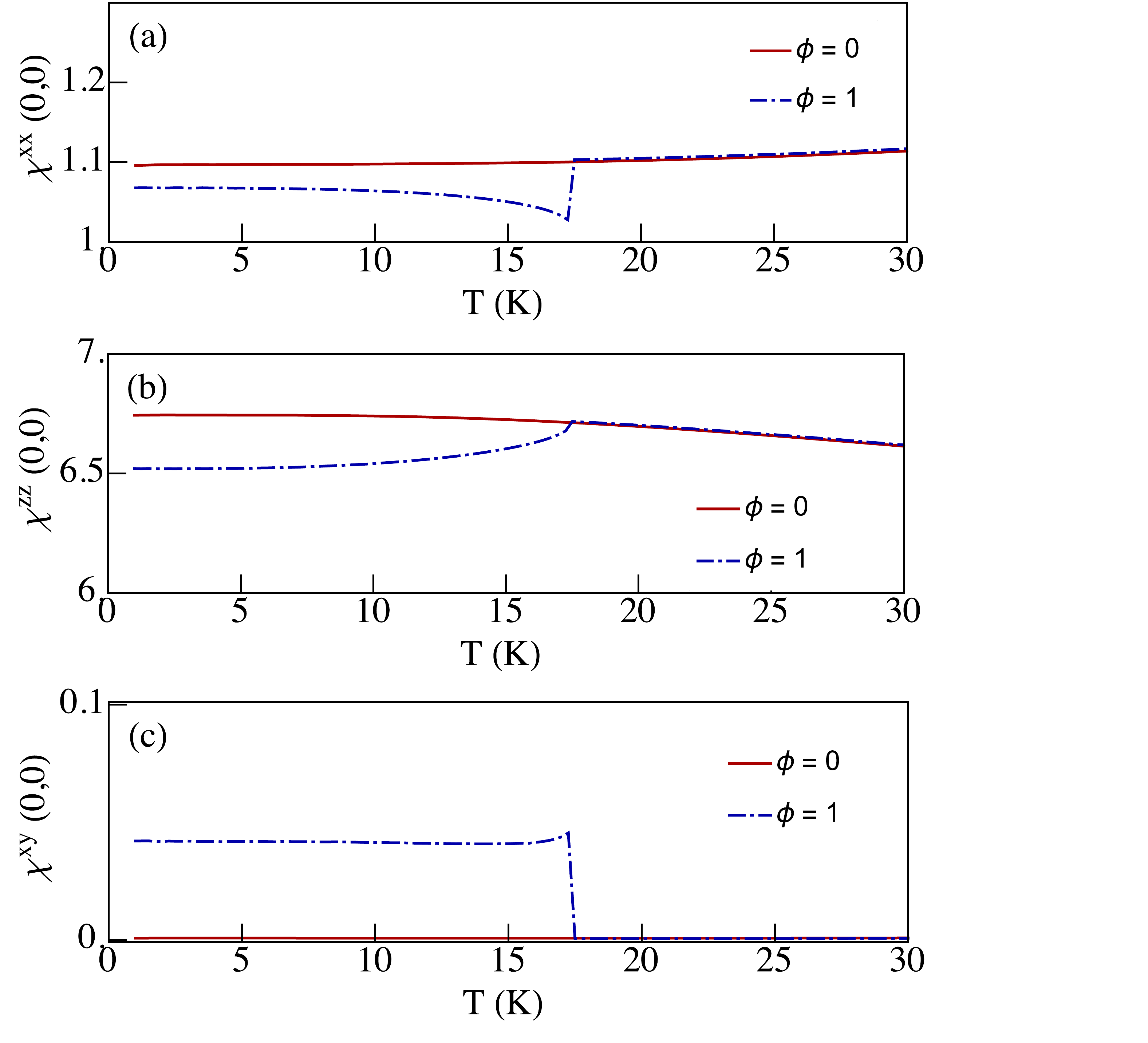}
\caption{
(Color online) 
Temperature evolution of the different tensor components of  static susceptibility $\hchi_0(q=0, \omega=0)$:
(a,b) $xx$ and $zz$ components exhibit suppression below $T_{HO}$. Note the scale difference due to larger magnetic moment 
matrix elements for z-direction (c) $xy$ component vanishes above $T_{HO}$ because of tetragonal symmetry and becomes finite below
due to fourfold symmetry breaking induced by $E_-(1,1)$ HO phase.
%\vspace{0.5cm}
}
\label{fig:Fig3}
\end{figure}
%%%%%%%%%%%%%%%%%%%%%%%%%%%%%%%%%%%%%%%%%%%%%%%%%%%%%%%%%%%%%%%%%%%%%
%
\be
\bl
J^\bq_x
=&
\sqrt{5}\bigl[
\cos2\theta(S^{\bq x}_{12}+S^{\bq x}_{21})+\sin2\theta(S^{\bq x}_{11}-S^{\bq x}_{22})\bigr],
%\non
\\
J^\bq_y
=&
\sqrt{5}\bigl[
\cos2\theta(S^{\bq y}_{12}+S^{\bq y}_{21})+\sin2\theta(S^{\bq y}_{11}-S^{\bq y}_{22})\bigr],
%\non
\\
J^\bq_z
=&
4\bigl[
\cos2\theta(S^{\bq z}_{11}-S^{\bq z}_{22})-\sin2\theta(S^{\bq z}_{12}+S^{\bq z}_{21})\bigr]
%\non
\\
&
+(S^{\bq z}_{11}+S^{\bq z}_{22}).
\label{eq:Jq}
\el
\ee
The physical magnetic moment is then obtained through $\bmu=(g\mu_B)\bJ$ with $g=6/7$.\\

The total angular momentum components in Eq.~(\ref{eq:Jq}) may also be expressed as bilinear forms of spinor operators $\psi^\dg_\bk=(\psi^\dg_{a\bk},\psi^\dg_{b\bk})$  of Eq.~(\ref{eq:NambuE}) by using Eq.~(\ref{eq:FTdef}). Then we have to restrict summation over \bk~to the first st BZ of the HO phase and obtain
\be
\bl
J_\lambda^\bq
&
=\fN\sum_\bk\psi^\dg_\bk\Gamma_\lambda\psi_{\bkq}
\\
&
=
\fN\sum_{\bk }
\left(
 \begin{array}{cc}
{\psi}^{\dg}_{a\bk} &{\psi}^\dg_{b\bk}
\end{array}
\right)
\left(
 \begin{array}{cc}
\Ga^0_\la& \Ga^1_\la\\
s_\la \Ga^1_\la & s_\la \Ga^0_\la
\end{array}
\right)
\left(
 \begin{array}{c}
{\psi}_{a\bkq} \\
{\psi}_{b\bkq}
\end{array}
\right).
\label{eq:momentq3}
\el
\ee
Here the signs $s_\la$ are given by $s_x=1$ and $s_y=s_z=-1$. The $\Ga^0_\la, \Ga^1_\la$ matrices are given  by
\be
\bl
\Ga^0_x
=&\tau_0\kappa_x\mu\cos 2\theta; \;\;\;\;\;\; \Ga^1_x=\tau_0\kappa_z\mu\sin 2\theta,
\\
\Ga^0_y
=&\tau_0(i\kappa_y)(i\mu)\cos 2\theta; \;\; \Ga^1_y=\tau_0\kappa_0(i\mu)\sin 2\theta,
\\
\Ga^0_z
=&\tau_0(2\cos 2\theta\kappa_0+\fs\kappa_z); \;\; \Ga^1_z=-\tau_0(i\kappa_y)2\sin 2\theta
.
\el
\ee
where $\kappa_x,\kappa_y,\kappa_z$ are the Pauli matrices and $\kappa_0$ the unit in orbital ($\al=1,2$) space while
$\tau_0$ is the unit in HO Nambu space. The explicit $\Ga^{0,1}_\la$ matrices are presented in Appendix \ref{sec:app2}.\\

Finally we have to transform to the spinors $\psi'_{\ga\bk}$ $(\ga=a,b)$ which diagonalize the 
Hamiltonian in the HO phase. This is achieved by the unitary transformation \cite{akbari:14} of Eq.~(\ref{eq:quasicoord})
resulting in
\bea
J_\la^\bq=\fN\sum_\bk\psi^{'\dg}_\bk\Gamma'_{\la\bk\bq}\psi'_{\bkq},
\label{eq:momentq4}
\eea
The transformed moment matrix elements which are now momentum dependent are given by $(\la=x,y,z)$
\be
\bl
\Gamma'_{\la\bk\bq}
&
=
\left(
 \begin{array}{cc}
U_{a\bk}\Ga^0_\la U^\dg_{a\bkq}& U_{a\bk}\Ga^1_\la U^\dg_{b\bkq}\\
s_\la U_{b\bk} \Ga^1_\la U^\dg_{a\bkq}& s_\la U_{b\bk}\Ga^0_\la U^\dg_{b\bkq}
\end{array}
\right)
\\
&
\equiv
\left(
 \begin{array}{cc}
\Ga^{'aa}_{\la\bk\bq}& \Ga^{'ab}_{\la\bk\bq}\\
\Ga^{'ba}_{\la\bk\bq} & \Ga^{'bb}_{\la\bk\bq}
\end{array}
\right)
.
\label{eq:gammapr}
\el
\ee
Here again $s_x=1, s_y=s_z=-1$.
The essential matrix elements for calculating the bare magnetic response functions $\chi^0_{\lambda\lambda'}(\bq,\om)$ in the HO phase are then contained in the matrices $\Ga^{\ga\ga'}_{\la\bk\bq}$   ($\ga,\ga'=a,b)$
where $\Gamma_\la^{0,1}$ are defined in Eq.~(\ref{eq:baregamma}) and $U_{\gamma\bk}$ is the unitary matrix
that diagonalizes the $4\times 4$ block Hamiltonian matrices $h_{\gamma\bk}$ of Eq.~(\ref{eq:hamblock}).
%
%%%%%%%%%%%%%%%%%%%%%%%%%%%%%%%%%%%%%%%%%%%%%%%%%%%%%%%%%%%%%%%%%%%%%
\begin{figure}
\includegraphics[width=0.95\linewidth,clip]{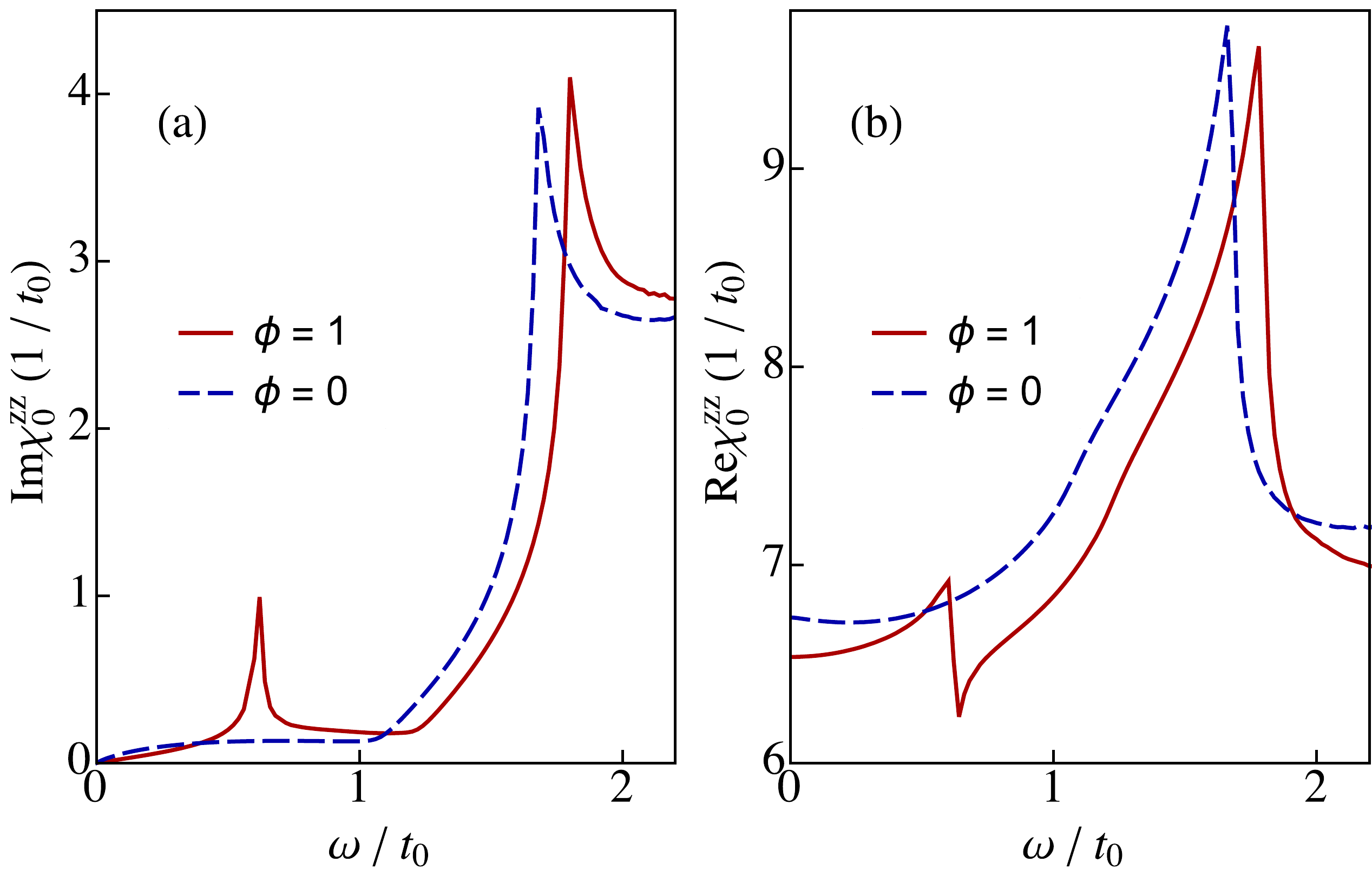}
\caption{
(Color online) 
Bare (noninteracting) dynamical susceptibility $\chi^{zz}_0(\bQ,\omega$ at the bct $Z$-point $(1,0,0)$ (r.l.u.) below $(\phi=1)$
and and above $(\phi=0)$ HO temperature: (a) is the imaginary part, and (b) shows the real part. The singular behaviour
in the low energy region is due to the HO gap opening and leads to the resonance appearance in Fig.~\ref{fig:Fig5}.
%\vspace{0.5cm}
}
\label{fig:Fig4}
\end{figure}
%%%%%%%%%%%%%%%%%%%%%%%%%%%%%%%%%%%%%%%%%%%%%%%%%%%%%%%%%%%%%%%%%%%%%
%

\section{The bare magnetic susceptibility in the HO phase}
\label{sec:baresus}
The bare dynamic magnetic susceptibility $(\la,\la'=x,y,z$ are cartesian matrix indices) is obtained from the Fourier transform of magnetic moment Green's function according to
\be
\bl
&
\chi_0^{\la\la'}(\bq,\omega)=
\\
&\hspace{0.2cm}
\frac{1}{N}\sum_{\bk,n,\bk' n'}
\frac{\bra n\bk|J_\la^\bq|n'\bk'\ket^*\bra n\bk|J_{\la'}^\bq|n'\bk'\ket}
{\vare_{n\bk}-\vare_{n'\bk'}-\omega -i\e}
(f_{n'\bk'}-f_{n\bk}).
\label{eq:sus1}
\el
\ee
Here $n=(\nu,\gamma)$ is a double index with $\nu=1-4$ denoting one of the four HO bands and $\gamma=a,b$ is the Kramers degeneracy. One can sum over the latter only in the matrix elements since $\vare_{n\bk}=\vare_{\nu\bk}$. To obtain the correct singular behaviour of $\chi_0^{\la\la'}(\bq,\omega)$ close to the HO gap it is essential to include the reconstruction of Bloch states by the HO parameter in the matrix elements of Eq.~(\ref{eq:sus1}).
They are nonvanishing when the sum of wave vectors corresponds to a reciprocal lattice vector $\bK$ of the simple tetragonal (st) BZ of the HO phase:
\be
\bl
\bra n\bk|J^\bq_\la| n'\bk'\ket
&=\bra n'\bk'|J^{-\bq}_\la| n\bk\ket^*
\\
&
=\sum_\bK\delta_{-\bk+\bq+\bk'+\bK,0} 
\;
m^{nn'}_{\la\la'}(\bk\bk')
.
\label{eq:susmat0}
\el
\ee
Inserting this into Eq.~(\ref{eq:sus1}) we obtain 
\be
\bl
&
\chi_0^{\la\la'}(\bq,\omega)
=\frac{1}{N}\sum_{\bk,\nu,\nu'}
\sum_{\ga\ga'\bK} 
\\
&\hspace{0.3cm}
 |m^{\nu\ga\nu'\ga'}_{\la\la'}(\bk,\bk-\bq-\bK)|^2
\frac{f_{\nu'\bk-\bq}-f_{\nu\bk}}
{\vare_{\nu\bk}-\vare_{\nu'\bk-\bq}-\omega -i\e}
,
\label{eq:sus2}
\el
\ee
where we used the Kramers degeneracy $\vare_{n\bk}=\vare_{\nu\bk}$ and periodicity of bands in the HO phase $\vare_{\nu\bk+\bK}=\vare_{\nu\bk}$. Explicitly, we have 
\be
\bl
&
\sum_{\ga\ga'\bK}|m_{\la\la'}^{\nu\ga\nu'\ga'}(\bk,\bk-\bq-\bK))|^2
=
\\
&
\sum_{\ga\ga'\bK}
\bra \nu\ga\bk|J_\la^\bq|\nu'\ga'\bk-\bq-\bK\ket^*\bra \nu\ga\bk|J_{\la'}^\bq|\nu'\ga'\bk-\bq-\bK\ket
.
\label{eq:susmat2}
\el
\ee
Changing the summation  in Eq.~(\ref{eq:sus2}) from $\bk$ into $-\bk$ and using inversion symmetry we can finally write
\be
\chi_0^{\la\la'}(\bq,\omega)=\frac{1}{N}\sum_{\bk,\nu,\nu'}
M_{\la\la'}^{\nu\nu'}(\bk\bq)
\frac{f_{\nu'\bkq}-f_{\nu\bk}}
{\vare_{\nu\bk}-\vare_{\nu'\bkq}-\omega -i\e}
.
\label{eq:sus3}
\ee
Using the explicit component representation of momentum operator of Eq.~(\ref{eq:momentq4}) according to
\be
J_\la^\bq=\fN\sum_{\bk\nu\nu'\ga\ga'}\psi^{'\dg}_{\nu\ga\bk}(\Gamma^{'\ga\ga'}_\la)^{\bk\bq}_{\nu\nu'}\psi'_{\nu'\ga'\bkq}
,
\label{eq:momentq5}
\ee
 in Eq.~(\ref{eq:susmat2})  we then obtain the matrix elements
\be
\bl
M^{\nu\nu'}_{\la\la'}(\bk\bq)
&=
\sum_{\ga\ga'\bK}|m_{\la\la'}^{\nu\ga\nu'\ga'}(\bk,\bkq+\bK))|^2
\\
&=
\frac{1}{N}\sum_{\ga\ga'\bK}
{(\Gamma^{'\ga\ga'}_\la)^{\bk\bq+\bK}_{\nu\nu'}}^*
(\Gamma^{'\ga\ga'}_{\la'})^{\bk\bq+\bK}_{\nu\nu'}
.
\label{eq:susmat3}
\el
\ee
For the HO phase with wave vector $\bQ=(\frac{2\pi}{a},0,0)$ only $\bK=0,\bQ$ are independent
because the $\Gamma'_{\la\bk\bq}$ matrices are periodic with 2\bQ~ and the energies with \bQ. Then the
sum over \bK~ contains only two terms leading the final explicit matrix elements:
\be
\bl
M^{\nu\nu'}_{\la\la'}(\bk\bq)
=
\sum_{\ga\ga'}\bigl[
&
{(\Gamma^{'\ga\ga'}_\la)^{\bk\bq}_{\nu\nu'}}^*
(\Gamma^{'\ga\ga'}_{\la'})^{\bk\bq}_{\nu\nu'}
\\&
+
{(\Gamma^{'\ga\ga'}_\la)^{\bk\bq+\bQ}_{\nu\nu'}}^*
(\Gamma^{'\ga\ga'}_{\la'})^{\bk\bq+\bQ}_{\nu\nu'}
\bigr],
\label{eq:susmat4}
\el
\ee
in terms of the transformed moment matrices of Eq.~(\ref{eq:gammapr}). 
They fulfil the periodicity $M^{\nu\nu'}_{\la\la'}(\bk\bq+\bQ)=M^{\nu\nu'}_{\la\la'}(\bk\bq)$ of the HO phase st lattice. Together with the periodicity
of the energy bands this means $\chi_0^{\la\la'}(\bq+\bQ,\omega)=\chi_0^{\la\la'}(\bq,\omega)$. For the diagonal cartesian susceptibility components
the above matrix elements simplify to 
\be
\bl
&
M^{\nu\nu'}_{\la\la}(\bk\bq)=
\\
&
\sum_\ga
\bigl[
|(\Gamma^{'\ga\ga}_\la)^{\bk\bq}_{\nu\nu'}|^2
+|(\Gamma^{'\ga\bar{\ga}}_\la)^{\bk\bq}_{\nu\nu'}|^2+( \bq \rightarrow \bq+\bQ)
%\\
%&
%|(\Gamma^{'\ga\ga}_\la)^{\bk\bq+\bQ}_{\nu\nu'}|^2
%+|(\Gamma^{'\ga\bar{\ga}}_\la)^{\bk\bq+\bQ}_{\nu\nu'}|^2
\bigr],
\label{eq:susmat4si}
\el
\ee
with the explicit $\Ga'_\la$ matrix elements 
\be
\bl
&
|(\Gamma^{'\ga\ga}_\la)^{\bk\bq}_{\nu\nu'}|^2=
|(U_{\ga\bk}\Ga^0_\la U^\dg_{\ga\bkq})_{\nu\nu'}|^2,
%\;\;\;\;
\\&
|(\Gamma^{'\ga\bar{\ga}}_\la)^{\bk\bq}_{\nu\nu'}|^2=
|(U_{\ga\bk}\Ga^1_\la U^\dg_{\bar{\ga}\bkq})_{\nu\nu'}|^2,
\label{eq:susmat5}
\el
\ee
and similar for wave vector \bq+\bQ. Here we used the convention $\bar{\gamma}=b,a$ for $\gamma=a,b$ respectively in the nondiagonal parts.
The Eqs.(\ref{eq:sus3},\ref{eq:susmat5}), together with HO bands $\vare_{\nu\bk}$, the HO transformation matrix $U_{\ga\bk}$ 
and the $\Gamma^{0,1}_\la$ matrices of Eq.~(\ref{eq:baregamma}) are the ingredients to calculate the diagonal susceptibility 
elements. The nondiagonal (xy-type) ones of general case in Eqs.~(\ref{eq:sus3},\ref{eq:susmat4}) can only appear in the (trigonal) E$_-(1,1)$ HO phase as induced elements and therefore will be quite small.
%
%%%%%%%%%%%%%%%%%%%%%%%%%%%%%%%%%%%%%%%%%%%%%%%%%%%%%%%%%%%%%%%%%%%%%
\begin{figure}
\includegraphics[width=\linewidth,clip]{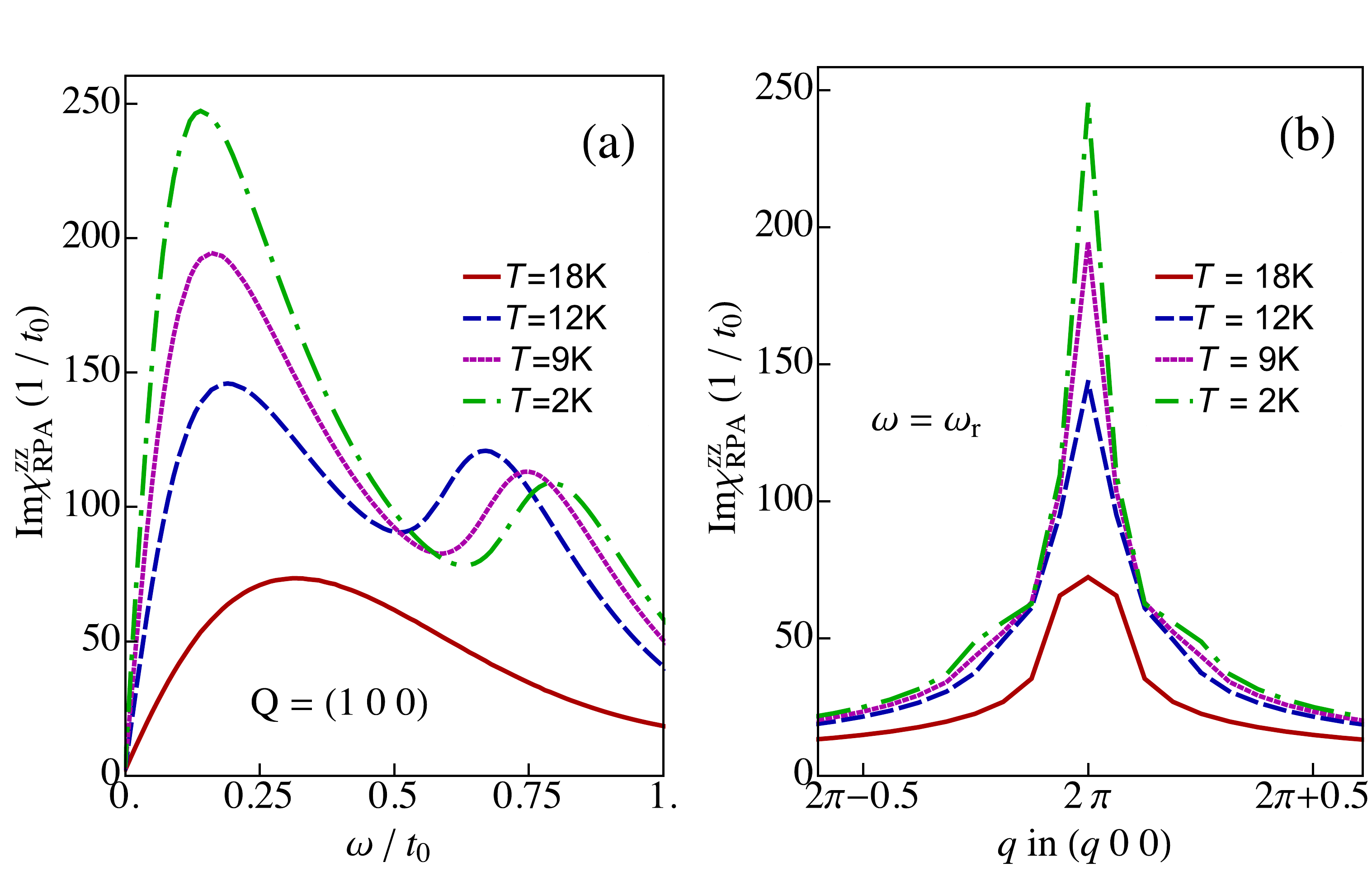}
\caption{
(Color online) 
(a) Collective magnetic $(zz,\perp)$ excitation spectrum: the imaginary part of the collective RPA susceptibility at the bct $Z$ point versus energy.
The position of the pronounced peak determines the (almost T- independent) resonance energy $\omega_r$  in the HO phase.
Here and in following figures $J_\perp(\bQ) = 0.156t_0 =1.04$ meV
(b) Evolution of  imaginary part of the RPA susceptibility at resonance energy $\omega_r(T)$ in q-scan around bct $Z$- point for different 
temperatures.
%\vspace{0.5cm}
}
\label{fig:Fig5}
\end{figure}
%%%%%%%%%%%%%%%%%%%%%%%%%%%%%%%%%%%%%%%%%%%%%%%%%%%%%%%%%%%%%%%%%%%%%
%

\section{The HO-RPA susceptibility and appearance of collective spin exciton resonance at $\bQ=(1,0,0)$}
\label{sec:RPA}

The bare magnetic response will be enhanced by effective exchange interactions between quasiparticles in the HO phase. This is also suggested
by the proximity of the AF phase that exists above a small critical pressure.  If the formation of heavy quasiparticles is described by the constrained (slave boson) mean-field approximation of an Anderson-lattice type model these quasiparticle interactions are caused by fluctations beyond the mean field level \cite{riseborough:92}.
However their momentum structure is too singular and therefore we use here a phenomenological form of the  exchange interactions
described by
\be
H_{ex}=-\sum_{\bq\la\la'}J_{\la\la'}(\bq)J_\la^\bq J_{\la'}^{-\bq},
\ee
 Generally, within the RPA approach the effective exchange function $\hJ(\bq)$ must have its maximum at the wave vector where the main resonance peak appears, which is the zone boundary (ordering) vector $\bQ = (1,0,0)$ (in this section r.l.u. units $\frac{2\pi}{a},\frac{2\pi}{c}$ of INS are used). The {\bf q} dependence around \bQ~may then be used for fitting to the range of the resonance dispersion. In \URU~the dispersion is well localized at \bQ~with only weak and steep dispersive features close to it \cite{bourdarot:14} as discussed below. The numerical results will show that this is dominated by the \bq- dependence of the bare susceptibility. Therefore we can use the simplest approximation of a \bq-independent $\hJ(\bq)=\hJ(\bQ)$ for the exchange. The numerical value of the interaction parameter $\hJ(\bQ)$ is then obtained by requiring the  peak position in the RPA spectrum to be close to the experimental resonance position.
In the tetragonal structure exchange function $\hJ(\bq)$  and susceptibility  $\hchi_0(\bq,\omega)$ are uniaxial tensors $\hJ(\bq)$ given by
\be
\bl
\hJ(\bq)
&=
\left(
 \begin{array}{ccc}
J_\parallel(\bq)& 0& 0 \\
0& J_\parallel(\bq) & 0 \\
0& 0 & J_\perp(\bq) \\
\end{array}
\right)
;\;%\;\;
\\
\hchi_0(\bq,\omega)
&=
\left(
 \begin{array}{ccc}
\chi_0^{xx}(\bq,\omega)& \chi_0^{xy}(\bq,\omega)& 0 \\
 \chi_0^{yx}(\bq,\omega)& \chi_0^{yy}(\bq,\omega)&0\\
0& 0 &  \chi_0^{\perp}(\bq,\omega)\\
\end{array}
\right).
\el
 \ee
The non-diagonal elements in the susceptibility matrix may appear for the $E_-(1,1)$ HO phase but not for the  $E_-(1,0)$,  $E_-(0,1)$ and disordered  phases. The collective RPA susceptibility then has the form
\be
\bl
\hchi(\bq,\omega)
&=[\hat{1}-\hchi_0(\bq,\omega)\hJ(\bq)]^{-1}\hchi_0(\bq,\omega)
\\
&=
\left(
 \begin{array}{cc}
\hchi^\parallel_{RPA}(\bq,\omega)& 0 \\
0& \chi^\perp_{RPA}(\bq,\omega)  \\
\end{array}
\right),
\label{eq:RPA1}
\el
\ee
where inversion leads to the final result
\be
\chi^\perp_{RPA}(\bq,\omega)=\Bigl[1-J_\perp(\bq)\chi_0^\perp(\bq,\omega) \Bigl]^{-1}\chi_0^\perp(\bq,\omega),
\label{eq:RPAperp}
\ee
\begin{widetext}
\be
\bl
\hchi^\parallel_{RPA}(\bq,\omega)&=&\frac{1}{D_\parallel(\bq,\omega)}
\left(
 \begin{array}{cc}
(1-J_\parallel\chi^{yy}_0)\chi_0^{xx}+J_\parallel\chi_0^{xy}\chi_0^{yx}&
(1-J_\parallel\chi^{yy}_0)\chi_0^{xy}+J_\parallel\chi_0^{yy}\chi_0^{xy} \\
(1-J_\parallel\chi^{xx}_0)\chi_0^{yx}+J_\parallel\chi_0^{xx}\chi_0^{yx}&
(1-J_\parallel\chi^{xx}_0)\chi_0^{yy}+J_\parallel\chi_0^{xy}\chi_0^{yx} 
\end{array}
\right)_{\bq,\omega}.
\label{eq:RPApar}
\el
\ee
\end{widetext}
Here the determinant is obtained from
\be
D_\parallel
\!
(\bq,\omega)
\!=\!
1-J_\parallel(\bq)(\chi_0^{xx}+\chi_0^{yy})+J_\parallel(\bq)^2
(\chi_0^{xx}\chi_0^{yy}-\chi_0^{xy}\chi_0^{yx}).
\ee
For vanishing non-diagonal ($xy,yx$) susceptibility elements  $\hchi_{RPA}^\parallel(\bq,\omega)$ is proportional to the unit matrix and $\chi_{RPA}^\parallel(\bq,\omega)$ is obtained from Eq.~(\ref{eq:RPAperp}) by replacing $\perp\rightarrow\parallel$.
%
%%%%%%%%%%%%%%%%%%%%%%%%%%%%%%%%%%%%%%%%%%%%%%%%%%%%%%%%%%%%%%%%%%%%%
\begin{figure}
\includegraphics[width=\linewidth,clip]{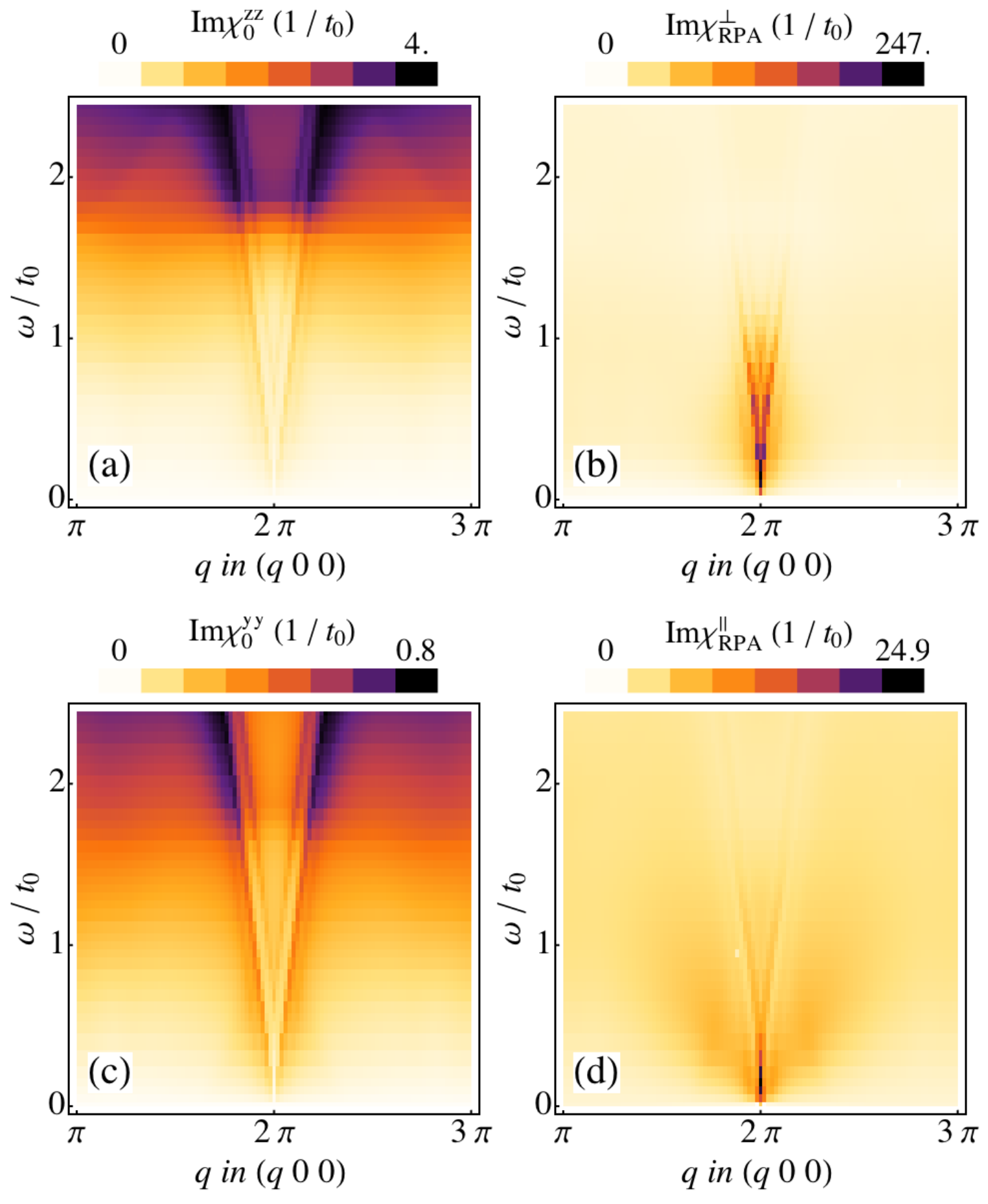}
\caption{
(Color online) 
(a)  Contour plot of the imaginary part of the bare susceptibility  $\chi_{0}^{zz}$ $\perp$ to tetragonal plane,  
(b)  the RPA susceptibility $\chi_{RPA}^{\perp}$    
(c) corresponding plots of the imaginary part of the bare and 
(d) RPA  spectrum  of ($yy,\parallel$) dynamical  susceptibility with a large $J_{\parallel}(\bQ)=0.91t_0=5.8$meV.
For istotropic $J_{\parallel}(\bQ)=J_\perp(\bQ)$ no resonance appears in this channel.
Here we set  $T=3$K,  in the  HO phase with $\phi=1$. The resulting resonance peak in (b),(d) is at $\omega_r=0.18$ meV. 
The dispersive V-shaped tails are remnants of the quasi-1D features of particle-hole continuum due to the nesting.
%\vspace{0.5cm}
}
\label{fig:Fig6}
\end{figure}
%%%%%%%%%%%%%%%%%%%%%%%%%%%%%%%%%%%%%%%%%%%%%%%%%%%%%%%%%%%%%%%%%%%%%
%
The dynamical structure function which is proportional to the INS scattering cross section \cite{bourdarot:10,villaume:08} of \URU~is given by $(\beta=(k_BT)^{-1})$:
\be
S(\bq,\omega)=\frac{1}{\pi}\frac{1}{1-e^{-\beta\omega}}
\sum_{\la\la'}(\delta_{\la\la'}-\hat{q}_\la\hat{q}_{\la'}) Im\chi_{\la\la'}(\bq,\omega).
\label{eq:struc1}
\ee
For small temperatures $(\beta\omega\gg 1)$ the Bose-factor $(1-e^{-\beta\omega})^{-1}\rightarrow 1$ and for large temperatures  $(\beta\omega\ll 1)$
$(1-e^{-\beta\omega})^{-1}\rightarrow (\beta\omega)^{-1}$.\\

In INS investigations the scattering vectors for the bct structure are conventionally indexed by those of the simple tetragonal structure \cite{butch:15}, i.e. by the cartesian components. Mainly the $[100]$ direction has been investigated \cite{bourdarot:10,villaume:08} sofar where  $\bq={\text q}\hat{\bq}_x$. Note that the in-plane zone boundary vectors $\bQ=(1,0,0)$ and $ (0,1,0)$ are equivalent to the Z-point vector $\bQ=(0,0,1)$ of the bct structure which is also the ordering vector of the antiferro- type HO. This is due to the fact that adjacent bct BZ's are shifted along [001] direction \cite{butch:15}.
For the [100] momentum direction we then obtain (i.f. $\beta\omega\ll 1$ or $T\rightarrow 0$):
\be
S(\bq,\omega)=\frac{1}{\pi}Im
\Big[
\chi^{\perp}_{RPA}(\bq,\omega)+\chi^{\parallel}_{RPA}(\bq,\omega)
\Big],
\label{eq:struc2}
\ee
which means that only moment fluctuations $\perp$ to the $[1,0,0]$ direction contribute to $S(\bq,\omega)$. 
Using the explicit RPA expressions of Eqs.~(\ref{eq:RPAperp},\ref{eq:RPApar}) we obtain:
\be
\bl
S(\bq,\omega)
&=\frac{1}{\pi}Im\Bigl[\frac
{\chi^\perp_0}
{1-J_\perp\chi^\perp_0}+
\\
&
\frac
{\chi_0^{yy}-J_\parallel(\chi_0^{xx}\chi_0^{yy}-\chi_0^{xy}\chi_0^{yx})}
{1-J_\parallel(\chi_0^{xx}+\chi_0^{yy})+J^2_\parallel(\chi_0^{xx}\chi_0^{yy}-\chi_0^{xy}\chi_0^{yx})}
\Bigr]_{\bq,\omega}.
\el
\label{eq:struc3}
\ee
We will later plot the two contributions separately for clarity. Because of the large uniaxial anisotropy of matrix elements (Appendix \ref{sec:app2}) the sum
will be determined by the $\perp (zz)$ contribution; for isotropic $J_\parallel=J_\perp$ only this channel develops the resonance peak.
It is also useful to consider two special cases that may be realized, depending on the type of HO, i.e., whether in-plane tetragonal symmetry is preserved or broken and in the latter case  whether $(1,1)$ or $(1,0)$- type orientation of the degenerate
$E_-$ order parameter is realized:\\

$i)$ Fourfold symmetry breaking but vanishing non diagonal elements:\\
This case corresponds to $E_-(10)$ or  $E_-(01)$  with  $\chi_{xx}\neq\chi_{yy}$ and $\chi_{xy}=\chi_{yx}=0$.
Then we have
\be
S(\bq,\omega)=\frac{1}{\pi}Im\Bigl[\frac
{\chi_0^\perp}
{1-J_\perp\chi_0^\perp}
+\frac
{\chi_0^{yy}}
{1-J_\parallel\chi_0^{yy}}
\Bigr]_{\bq,\omega}.
\ee
If the fourfold symmetry breaking is absent $\chi_0^{yy}=\chi_0^{xx}=\chi_0^\parallel$.
Then the above expression is formally the same as for the disordered phase $(\phi=0)$ above T$_{HO}$.\\
%
%%%%%%%%%%%%%%%%%%%%%%%%%%%%%%%%%%%%%%%%%%%%%%%%%%%%%%%%%%%%%%%%%%%%%
\begin{SCfigure*}
 \centering
 \caption{ 
(Color online) 
3D plot of the imaginary part of the perpendicular component of (a) bare (noninteracting ) $\chi^{zz}_0$ and (b) RPA  susceptibility $\chi_{RPA}^{\perp}$ in hidden order phase with  with $\phi=1$ and $T=3$K (same as Fig.\ref{fig:Fig6}b). In the latter main resonance peak and V-shaped dispersive features at higher energies can be identified.
\vspace{0.5cm}
 }
 \includegraphics[width=0.73\textwidth]%
{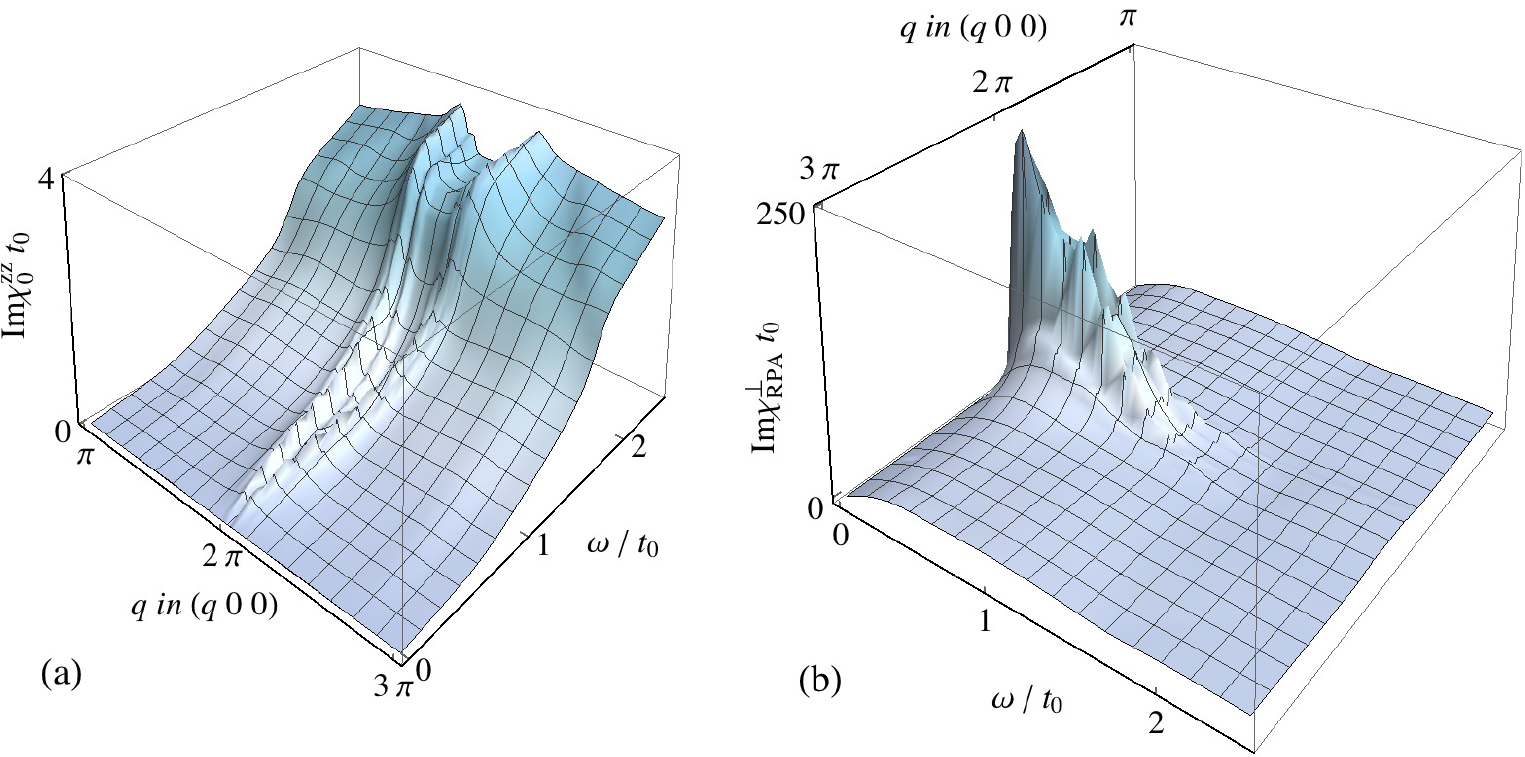}% picture filename
\label{fig:Fig7}
\end{SCfigure*}
%%%%%%%%%%%%%%%%%%%%%%%%%%%%%%%%%%%%%%%%%%%%%%%%%%%%%%%%%%%%%%%%%%%%%
%
$ii)$ Fourfold symmetry breaking through finite and equal non-diagonal elements:\\
This case corresponds to $E_-(11)$ or $E_-(1\bar{1})$ with  
$\chi_{xx}=\chi_{yy}\equiv\chi^\parallel_0$ and $\chi_{xy}=\chi_{yx}\equiv\tilde{\chi}_0$ leading to 
\be
S(\bq,\omega)=\frac{1}{\pi}Im\Bigl[\frac
{\chi_0^\perp}
{1-J_\perp\chi_0^\perp}
+\frac
{\chi^\parallel_0-J_\parallel(\chi^{\parallel 2}_0-\tilde{\chi}_0^2)}
{(1-J_\parallel\chi^\parallel_0)^2-J_\parallel^2\tilde{\chi}^2_0}
\Bigr]_{\bq,\omega}.
\ee
If the non-diagonal element $\tilde{\chi}_0$ induced by  $E_-(11)$ HO is negligible this reduces again to the simple RPA expression
\be
S(\bq,\omega)=\frac{1}{\pi}Im\Bigl[\frac
{\chi_0^\perp}
{1-J_\perp\chi_0^\perp}
+\frac
{\chi_0^{\parallel}}
{1-J_\parallel\chi_0^{\parallel}}
\Bigr]_{\bq,\omega},
\ee
which is formally identical to the disordered phase but $\chi_0^{\perp,\parallel}(\bq,\omega)$ now contain the effect of HO  band reconstruction. For nonzero temperatures the above expressions have to be multiplied
by $(1-e^{-\beta\omega})^{-1}$. 

\section{Discussion of numerical results for the magnetic excitation spectrum}
\label{sec:discussion}

Now we discuss the numerical results for the spin dynamics in the two-orbital model of \URU. First we focus on the bare noninteracting susceptiblility. The static, homogeneous ($\bq=0$) tensor components are shown in Fig.~\ref{fig:Fig3} as function of temperature. The large uniaxial anisotropy $\chi_{zz}/\chi_{xx}\simeq 6.1$ which is due to the CEF is comparable to the experimental one \cite{palstra:85}. It is used to determine the CEF mixing angle $\theta\simeq0.345\pi$  in Eq.~(\ref{eq:CEFtrans}).
Below $T_{HO}$  $\chi_{xx}=\chi_{yy}$ as well as $\chi_{zz}$ show a considerable depression caused by  the HO gap opening. 
Due to different selection rules for $J_x$ and $J_z$- operators for band states their behaviour around $T_{HO}$ is distinct. On the other hand the non-diagonal in-plane susceptibility $\chi_{xy}$ which has to vanish above $T_{HO}$ becomes finite in the HO phase. This is the reason for the appearance of twofold torque-oscillations in the HO-phase \cite{thalmeier:11,okazaki:11}.\\

The bare dynamical susceptibility (the Lindhard function) at the Z-point is shown in Fig.~\ref{fig:Fig4}a. The HO gap opening produces singularities in the response around the gap threshold $\omega \simeq \Delta_{HO}=\phi/\sqrt{2}$. These singularities are responsible for the resonance appearance in the collective response function. To obtain them one must use the reconstructed eigenstates and matrix elements of Eqs.~(\ref{eq:gammapr},\ref{eq:susmat4}).
Note there is an additional peak behaviour at higher energies $\omega\simeq 2t_0$ which is connected with the van Hove singularities of the band structure itself and therefore persists above $T_{HO}$ when $\phi=0$.\\

The imaginary part of the collective RPA susceptibility in the  $\perp (zz)$ channel, i.e., the magnetic excitation spectrum of the interacting itinerant 5f moments is shown in Fig.~\ref{fig:Fig5} for the
experimental [100] direction. The frequency dependence in Fig.~\ref{fig:Fig5}a at the bct Z point (100) shows the evolution of the spin resonance excitation out of the normal state spin fluctuation continuum (T=18 K)  when the temperature drops below $T_{HO}=17.5$ K. The interaction parameters (see caption) in the model are chosen such that the enhancement of peak intensity $I(T=0)/ I(T_{HO})\simeq 4$  relative to the disordered state and the position of the resonance at $\omega_r$ are approximately reproduced. The latter is at $\omega_r\simeq 0.18t_0 =1.2$ meV ($t_0=6.7$ meV) or $\omega_r(T=0)/\Delta_{HO}\simeq 0.29 $. This is  in reasonable agreement with the experimental results $\omega_r(T=0)/\Delta_{HO}\simeq 0.41 $ where $\omega_r=1.7$ meV \cite{bourdarot:10} and $\Delta_{HO}= 4.1$ meV  \cite{aynajian:10}.  There is almost no temperature dependence of the resonance position once it exists. This is also in agreement with experiment \cite{bourdarot:10,bourdarot:14}. On the other hand no clear indication of spin gap behaviour at the lowest $\omega$ precisely at the resonance vector \bQ~is obtained. There is also a smaller side hump at larger energy due to the higher energy singularity in $\chi_0(\bQ,\omega)$ (Fig.~\ref{fig:Fig4}). Alternatively we show a q-scan along $(q,0,0)$ direction for constant $\omega=\omega_r(T)$  at various temperatures. Again the resonance peak clearly grows below $T_{HO}$.\\

These results may also be demonstrated in contour plots of the magnetic spectrum in the $q,\omega$ plane (Fig.~\ref{fig:Fig6}). In (a) the spectrum of the bare dynamical susceptibility $\chi_0^{zz}$ is shown which exhibits the HO gap at $\omega/t_0 \geq 1$ and a V- shaped structure emerging from the bct Z-point ($q=\frac{2\pi}{c}$). Turning on the quasiparticle interaction leads to the spectrum of the collective RPA susceptibility $\chi^\perp_{RPA}$ shown in (b). Most of the magnetic intensity is now concentrated at the bct Z point resonance energy $\omega_r$ (c.f. Fig.~\ref{fig:Fig5}b). But still an indication of the V-shaped dispersion is visible. Qualitatively it agrees with the experimentally observed dispersion \cite{bourdarot:14} around Z. One may question about its origin. In this model the HO is driven by nesting between the two FS sheets which contains states with high angular momentum component. The nesting means there is a quasi-1D contribution to the spectrum of the bare susceptbility. In such quasi 1D-situation for low energies and small momenta with respect to nesting vector it consists of a very narrow particle-hole continuum with a dispersion $v^\parallel_Fq'$ where $q'=q-\frac{2\pi}{c}$ and $v^\parallel_F$ is the Fermi velocity in x-direction. The V-shaped ridges in (a,b) are their image. The spectrum of bare and RPA in-plane susceptibility $(yy,\parallel)$ shown in (c,d) respectively depict a qualitatively similar behaviour as described above, although the scale is much smaller due to the uniaxial matrix element anisotropy origninating in the CEF effect. Therefore its contribution to $S(\bq,\omega)$ in Eqs.~(\ref{eq:struc2},\ref{eq:struc3}) is less important.
Finally, the bare and RPA spectrum of magnetic excitations ($zz,\perp$ channnel) are presented in Fig.~\ref{fig:Fig7}a,b , respectively in a 3D perspective plot where the described features of resonance formation and attached ridge-like dispersion can be seen even more clearly.

\section{Spin gap formation and influence on NMR relaxation}
\label{sec:NMR}

\begin{figure}
\includegraphics[width=0.8\linewidth,clip]{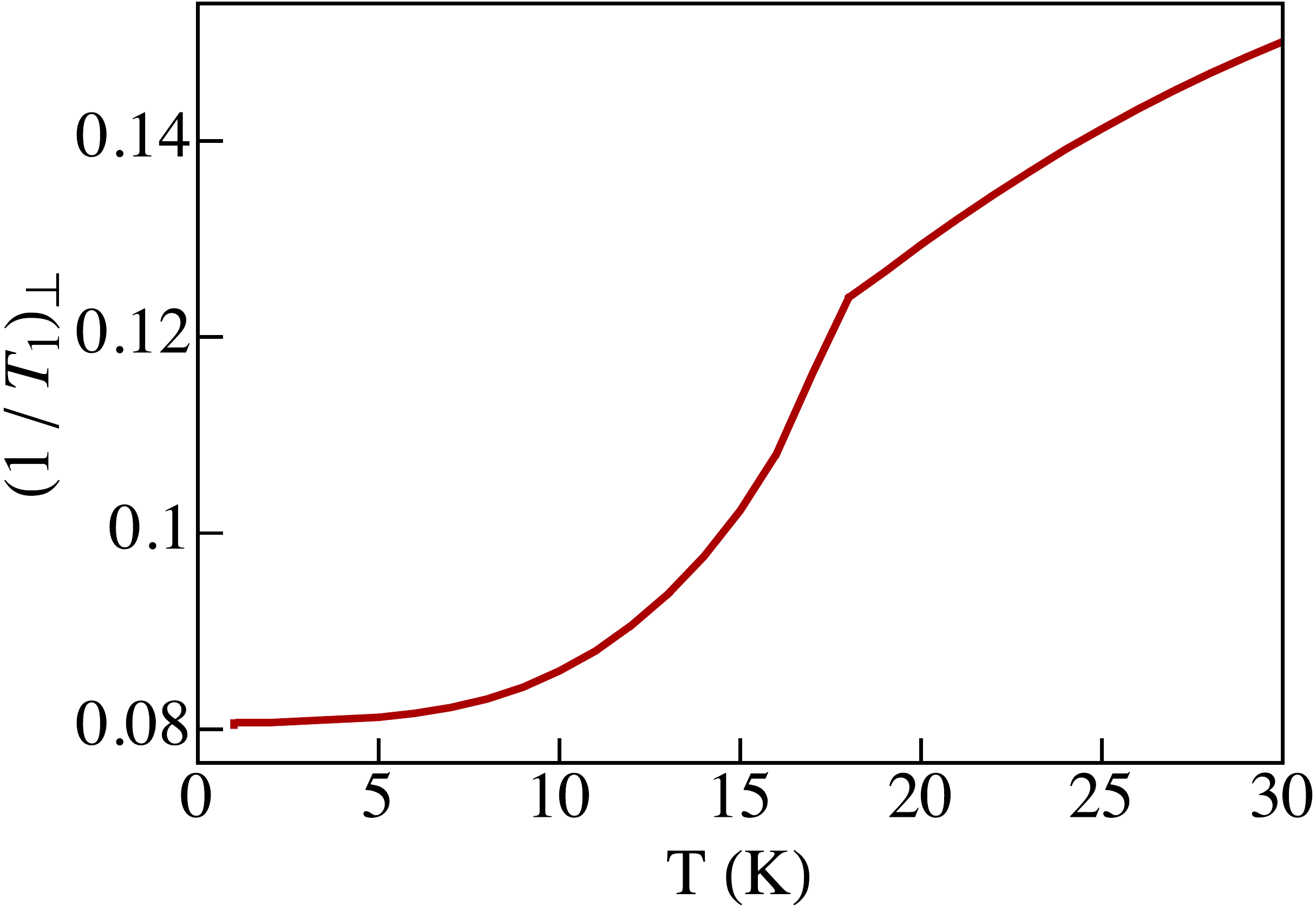}
\caption{
(Color online) 
Temperature dependency of  the normalised NMR relaxation rate,  $(T_1)_\perp$, at
NMR resonance frequency~$\omega_0\ll \omega_r$. Relaxation rate $(T_1)_\parallel$ has similar T-dependence but is larger due to
uniaxial anisotropy of magnetic matrix elements.
%\vspace{0.5cm}
}
\label{fig:Fig8}
\end{figure}

It is known from many examples that there is a dual relationship between spin resonance formation around the gap threshold (in this case HO gap) and 
a spin gap formation at low energies \cite{thalmeier:15} which are both seen in INS experiments. The spin gap formation also directly influences the NMR relaxation rate which is determined by low energy spin fluctuations. If the latter open a spin gap the relaxation rate should strongly decrease. This was indeed found in the NMR experiments \cite{kohori:96,emi:15} in \URU~below the hidden order transition T$_\text{HO}$. The normalized NMR relaxation rate at NMR resonance frequency $\omega_0$ is given via the dynamical RPA susceptibility by the relations
\be
\bl
\Bigl(\frac{1}{\hat{T}_1}\Bigr)_\perp =& t_0^2\sum_\bq\frac{Im\chi_\parallel(\bq,\omega_0)}{\omega_0},
%\non
\\
\Bigl(\frac{1}{\hat{T}_1}\Bigr)_\parallel =& t_0^2\sum_\bq
\frac{1}{2}\bigl[\frac{Im\chi_\parallel(\bq,\omega_0)}{\omega_0}+\frac{Im\chi_\perp(\bq,\omega_0)}{\omega_0}\bigr],
\el
\ee
where $t_0$ is the energy scale of HO bands used previously. The relaxation rates are normalized to the Korringa rate  $(1/T_1)_0=2\ga_n^2A_{hf}^2k_BT/(\ga_e\hbar)^2t_0^2$ where $A_{hf}$ is assumed as a \bq- and axis- $(\parallel,\perp)$ independent hyperfine coupling. The $\ga_n,\ga_e$ are nuclear and electronic gyromagnetic ratios. To comply with previous convention $(\parallel,\perp)$ denotes directions parallel and perpendicular to the tetragonal ab-plane (note this is opposite to conventions in Ref.~\onlinecite{kohori:96,emi:15}). The calculation of relaxation rate requires the \bq-dependence of dynamical susceptibility at resonance frequency $\omega_0$ in the whole st HO Brillouin zone, not just along the \bQ-direction. The results are shown in Fig.~\ref{fig:Fig8} for a frequency $\omega_0\ll \Delta_{HO}$, i.e., much smaller than the HO gap. Below $T_{HO}$ a pronounced reduction is observed  which shows that an overall low energy spin-gap is developing in the HO phase although it is not localized in momentum space around the resonance vector \bQ~because the HO gapping is rather incomplete.
Qualitatively it is very similar to experimental results for $^{29}$Si-NMR in Ref.~\onlinecite{emi:15}. We only present the $(1/T_1)_\perp$ rate. The corresponding $(1/T_1)_\parallel$ rate shows almost identical behaviour except for an enhancement factor due to the susceptibility anisotropy which is close to the experimental enhancement $(1/T_1)_\parallel/(1/T_1)_\perp$ of about $4.5$ \cite{emi:15}.
%
%%%%%%%%%%%%%%%%%%%%%%%%%%%%%%%%%%%%%%%%%%%%%%%%%%%%%%%%%%%%%%%%%%%%%
%%%%%%%%%%%%%%%%%%%%%%%%%%%%%%%%%%%%%%%%%%%%%%%%%%%%%%%%%%%%%%%%%%%%%
%

\section{Summary and conclusion}
\label{sec:summary}

In this work we have given a theoretical analysis of the spin-resonance phenomenon in HO phase of  \URU. The physical properties of this collective excitation has been well investigated before by INS experiments \cite{bourdarot:10,bourdarot:14}. It appears inside the HO charge gap $\Delta_{HO}\simeq 4.1$ meV  at an energy $\omega_r=1.7$ meV at the commensurate wave vector $\bQ=(\frac{2\pi}{a},0,0)$ of the bct Z-point and it exhibits a V-shaped upward dispersion. Such spin-resonance excitations are frequently observed inside the quasiparticle gap of  unconventional superconductors like cuprates, pnictides and heavy fermion metals. More rarely they are found within the hybridization gap of Kondo insulators or the gap of hidden order compounds as in the present case.

For a semi-quantitative explanation of this interesting many body effect in \URU~ we used a previously investigated \cite{rau:12,akbari:14} two-orbital model of 5f electrons resulting in two FS sheets with states that contain high angular momentum components. The hidden order is then taken as the rank-5 doubly degenerate $E_-$ representation which agrees with all observed broken symmetries. Within a mean field HO theory the quasiparticle states in the HO phase are reconstructed and a HO gap is opening leading to a vanishing of large parts of the Fermi surface and associated DOS suppression of heavy charge carriers. The reconstructed states lead to singular behaviour of the bare magnetic response function. This enables the RPA response function of interacting quasiparticles to develop a pronounced resonance peak below the HO gap at the nesting vector \bQ~ with a temperature evolution below $T_{HO}$ that is similar to the experimental one. In addition V-shaped dispersive features appear at higher energies which have also been observed in INS. Although there is no pronounced spin gap evolution at \bQ~itself the momentum integrated low energy response seen in NMR develops such a suppression which is seen in the decrease of the NMR relaxation rate below $T_{HO}$.

Embedded in the HO phase an unconventional SC gap of presumably chiral d-wave nature appears \cite{kasahara:09,yano:08,schemm:15}. Its effect on QPI was studied previously \cite{akbari:14}. Because $T_c\ll T_{HO}$ by an order of magnitude the same is true for the superconducting gap with respect to HO gap. Therefore we were not able to see a significant change of the magnetic RPA response function when including the superconducting gap. This is also in agreement with INS findings \cite{hassinger:09}.

%%%%%%%%%%%%%%%%%%%%%%%%%%%%%%%%%%%%%%%%%%%%%%%%%%%%%%%%%%%%%%%%%%%%%%%%%%%%%%%%%%%%%%%%%
\section*{Acknowledgments}
 We are grateful to the Max Planck Institute for the Physics of Complex Systems (MPI-PKS) for the use of computer facilities.
A.A. acknowledges support by  Max Planck POSTECH / KOREA Research Initiative (No. 2011-0031558) programs through NRF funded by MSIP of Korea. 

%%%%%%%%%%%%%%%%%%%%%%%%%%%%%%%%%%%%%%%%%%%%%%%%%%%%%%%%%%%%%%%%%%%%%%%%%%%%%%%%%%%%%%%%%

%\newpage

\appendix

\section{Kinetic energy coefficients and para phase band structure}
\label{sec:app1}

For completeness and convenience  we recapitulate the effective 5f- two band model for \URU~that is adopted from Ref.~\onlinecite{rau:12} and also used previously in QPI calculations \cite{akbari:14}. The kinetic terms in  Eq.~(\ref{eq:bands}) are defined by the intra-orbital energies ($\al =1,2$ is the orbital index):
\be
\bl
A_{\al\bk}=&A_{\al\bk}^z+A_{\al\bk}^\perp +\frac{1}{2}sign(\al)\Delta_{12} ,
%\non
\\
A_{\al\bk}^\perp=&
2t'_\al(\cos ak_x +\cos ak_y) 
%\non\\&
+4t''_\al\cos ak_x\cos ak_y-\epsilon_0,
\\
A_{\al\bk}^z=&8t_\al\cos\fa k_x\cos\fa k_y\cos\fc k_z,
%\non
%\non
\el
\ee
where we defined $sign(\al)=(-1)^{\al-1}$ and the inter-orbital hopping energy 
\be
\bl
D_\bk
=
&
D'_\bk+iD''_\bk 
\\
=
&
t_{12}\bigl[\sin\fa(k_x+k_y)
-i\sin\fa(k_x-k_y)\bigr]\sin\fc k_z,
\label{eq:kinetic}
\el
\ee
To reproduce a realistic Fermi surface model with nesting electron- and hole- like pockets 
around the $\Gamma$ and $Z$ points of the bct Brillouin zone we use the following parameters \cite{rau:12}:
The orbital energy splitting is  $\Delta_{12}=3.5$ or $\Delta\equiv 0.5 \Delta_{12}=1.75$.
The nearest neighbor hoppings are  $t_1=t_2\equiv t =-0.3$, meaning orbital-independent $A_{\al\bk}^z=A_{\bk}^z$. Hopping elements to next and second nearest neighbors are $t'_1=-0.87$, $t'_2=0.0, t''_1=0.375, t''_2=0.25$, respectively and the average orbital energy is $-\epsilon_0 = 0.5$. The inter-orbital hopping is  $|t_{12}|=0.7$. All energies are given in units of $t_0=6.7$ meV corresponding to a total band width  \cite{akbari:14} $W_{qp}\simeq 12t_0=80$ meV obtained from tunneling results \cite{aynajian:10,park:12}. For the computation of quasiparticle bands in the HO phase it is useful to introduce (anti-) symmetrized
quantities:
\be
\bl
A_\bk^\perp=&\frac{1}{2}(A_{1\bk}^\perp+A_{2\bk}^\perp)
\\
=&2t'(\cos ak_x +\cos ak_y) +4t''\cos ak_x\cos ak_y -\epsilon_0,
\\
\De_\bk^\perp=&\Delta+\frac{1}{2}(A_{1\bk}^\perp-A_{2\bk}^\perp)
\\
=&
\Delta+2\de'(\cos ak_x +\cos ak_y) +4\de''\cos ak_x\cos ak_y.
\label{eq:kinsymm}
\el
\ee
Here  $t'=\frac{1}{2}(t'_1+t'_2)$,  $t''=\frac{1}{2}(t''_1+t''_2)$ and  
$\de'=\frac{1}{2}(t'_1-t'_2)$,  $\de''=\frac{1}{2}(t''_1-t''_2)$.
The auxiliary functions above have the following symmetry under translation by the ordering vector \bQ:
$A_{\al\bk+\bQ}^\perp=A_{\al\bk}^\perp$ implying also $A_{\bk+\bQ}^\perp=A_{\bk}^\perp$ and  
$\De_{\bk+\bQ}^\perp=\De_{\bk}^\perp$. On the  other hand $A_{\al\bk+\bQ}^z=-A_{\al\bk}^z$ and
$D_{\bk+\bQ}=-D_\bk$.\\

\section{Total angular momentum and susceptibility matrix elements for special cases}
\label{sec:app2}

Here we give the explicit form of  the magnetic moment matrices in the basis of free single ion $|j,M\rangle$ states in the 
$|j,\pm\frac{3}{2}\rangle$,  $|j,\pm\frac{5}{2}\rangle$ subspace that are needed to construct the total angular momentum operators in Eq.~(\ref{eq:momentloc1}).
$\hat{J}_\lambda$ $(\lambda =x,y,z$) can be written as (index $M$, $M'$ in the order $\fu,\fl,-\fl,\fu$), defining
$\mu=\frac{1}{2}\sqrt{5}=1.12$:
\be
\bl
&
\hat{J}_x=
\left(
 \begin{array}{cccc}
0& \mu & 0 & 0 \\
\mu& 0 & 0 & 0 \\
0& 0 & 0 & \mu \\
0& 0 & \mu & 0
\end{array}
\right)
;\;\;
\hat{J}_y=
\left(
 \begin{array}{cccc}
0& -i\mu & 0 & 0 \\
i\mu& 0 & 0 & 0 \\
0& 0 & 0 & -i\mu \\
0& 0 & i\mu & 0
\end{array}
\right)
;
%\;\;
\\&
\hspace{2cm}
\hat{J}_z=
\left(
 \begin{array}{cccc}
\fu& 0 & 0 & 0 \\
0 & \fl & 0 & 0 \\
0& 0 & -\fl & 0 \\
0& 0 & 0 & -\fu
\end{array}
\right).
\label{eq:momentmat}
\el
\ee
%\\
The susceptibility matrix elements $M^{\nu\nu'}_{\la\la'}$ for the normal state $(\phi=0)$ are defined in Eq.~(\ref{eq:susmat3}) via the unitary transforms of
the $\Ga_\la^{0,1}$ matrices in the space of four-dimensional $\psi_{\ga,\bk}, (\ga=a,b)$ spinors in Eq.~(\ref{eq:gammapr}).
Explicitly, in each of $\psi_{a\bk}$ or  $\psi_{b\bk}$ subspace they are given by ($\theta=$ CEF mixing angle):\\
\begin{widetext}
\be
\bl
&
\Ga^0_x=
\left(
 \begin{array}{cccc}
0& \mu\cos 2\theta & 0 & 0 \\
\mu\cos 2\theta & 0 & 0 & 0 \\
0& 0 & 0 & \mu\cos 2\theta \\
0& 0 & \mu\cos 2\theta & 0
\end{array}
\right)
;\;\;
%\\&
\Ga^1_x=
\left(
 \begin{array}{cccc}
\mu\sin 2\theta&0 & 0 & 0 \\
0& -\mu\sin 2\theta & 0 & 0 \\
0& 0 & \mu\sin 2\theta & 0 \\
0& 0 & 0 & -\mu\sin 2\theta
\end{array}
\right)%\non
\el
\ee
\be
\bl
%\\
&
\Ga^0_y=
\left(
 \begin{array}{cccc}
0& i\mu\cos 2\theta & 0 & 0 \\
-i\mu\cos 2\theta & 0 & 0 & 0 \\
0& 0 & 0 & i\mu\cos 2\theta  \\
0& 0 &- i\mu\cos 2\theta  & 0
\end{array}
\right)
;
\;\;
%\\&
\Ga^1_y=
\left(
 \begin{array}{cccc}
i\mu\sin 2\theta&0 & 0 & 0 \\
0& i\mu\sin 2\theta & 0 & 0 \\
0& 0 & i\mu\sin 2\theta & 0 \\
0& 0 & 0 & i\mu\sin 2\theta
\end{array}
\right)
\label{eq:baregamma}
\el
\ee
\be
\bl
%\\
&
\Ga^0_z=
\left(
 \begin{array}{cccc}
2\cos 2\theta+\fs& 0& 0 & 0 \\
0& 2\cos 2\theta-\fs & 0 & 0 \\
0& 0 & 2\cos 2\theta+\fs & 0 \\
0& 0 & 0 & 2\cos 2\theta-\fs
\end{array}
\right)
;
\;\;
%\\&
\Ga^1_z=
\left(
 \begin{array}{cccc}
0&-2\sin 2\theta & 0 & 0 \\
2\sin 2\theta& 0& 0 & 0 \\
0& 0 & 0 & -2\sin 2\theta  \\
0& 0 & 2\sin 2\theta  & 0
\end{array}
\right) % \non
\el
\ee
\end{widetext}
Now we derive the susceptibility matrix elements $M^{\nu\nu'}_{\la\la'}$ for the special case $D_\bk =0$ and without HO ($\phi=0$). Then Hamilton block matrices of Eq.~(\ref{eq:hamblock}) are already diagonal and therefore trivially $U_{a,b}=1$. The  $\Gamma'_{\la\bk\bq}$ are therefore momentum independent and equal to the bare $\Ga_\la$ matrices. Then the diagonal susceptibiltiy matrix elements of Eq.~(\ref{eq:susmat5}) reduce to the momentum independent expressions given by
\bea
M^{\nu\nu'}_{\la\la}=4[|(\Ga_\la^0)_{\nu\nu'}|^2+|(\Ga_\la^1)_{\nu\nu'}|^2].
\eea
Written in concise $4\times 4$ matrix form $\hat{M}_{\la\la}$ in $(\nu\nu')$ indices we get for the three cartesian directions:
\be
\bl
&\hat{M}_{xx}=\hat{M}_{yy}=4\mu^2\tau_0(\sin^22\theta\ka_0+\cos^22\theta\ka_x),
\\
&\hat{M}_{zz}=4\tau_0\bigl[(4\cos^22\theta+\frac{1}{4})\ka_0+2\cos2\theta\ka_z+4\sin^22\theta\ka_x\bigr],
\el
\ee
where $\tau_0,\ka_0$ are the units in Nambu and orbital space, respectively and $\ka_x,\ka_z$ are orbital Pauli matrices. In explicit matrix representation in $\psi_{\ga\bk}$ -spaces (with order $1\bk, 2\bk, 1\bk+\bQ, 2\bk+\bQ; 1,2=$ orbital index) we have
\begin{widetext}
\be
\bl
\hat{M}_{xx}=\hat{M}_{yy}=4\mu^2
\left(
 \begin{array}{cccc}
\sin^22\theta& \cos^22\theta & 0 & 0 \\
\cos^22\theta & \sin^22\theta & 0 & 0 \\
0& 0 &\sin^22\theta & \cos^22\theta \\
0& 0 & \cos^22\theta & \sin^22\theta
\end{array}
\right),
\label{eq:matxx}
\el
\ee
\be
\bl
\hat{M}_{zz}=4
\left(
 \begin{array}{cccc}
(2\cos 2\theta+\fs)^2& 4\sin^22\theta& 0 & 0 \\
4\sin^22\theta& (2\cos 2\theta-\fs)^2 & 0 & 0 \\
0& 0 & (2\cos 2\theta+\fs)^2 &4\sin^22\theta  \\
0& 0 & 4\sin^22\theta & (2\cos 2\theta-\fs)^2
\end{array}
\right).
\label{eq:matzz}
\el
\ee
\end{widetext}
The sum of all susceptibility matrix elements $\mu_\la=\sum_{\nu\nu'}M^{\nu\nu'}_{\la\la}$ is a measure of the effective moment in direction $\la$. They are independent of the CEF mixing angle $\theta$ because the latter causes just a rotation in the j=5/2 subspace. We obtain $\mu_z=64$ and
$\mu_x=\mu_y=8\mu^2=10.03$ leading to an anisotropy $\mu_z/\mu_x$=6.4 independent of $\theta$. This is approximately the anisotropy ratio of static susceptibility. The blocks with zeroes in $\hat{M}_{\la\la}$ appear because for $\phi=0$ no mixing of states with momenta \bk~and \bk+\bQ~is present. Finally we note that for $\phi=0$ we also have $\hat{M}_{\la\la'}\equiv 0$ $(\la\neq\la')$ because of preserved tetragonal symmetry. 
Therefore the $\hchi$ tensor in the disordered phase is uniaxial with $\chi^0_{xx}=\chi^0_{yy}\neq\chi^0_{zz}$.

\bibliography{References}

\end{document}